\documentclass[11pt]{article}

\usepackage[preprint]{acl}
\usepackage{times}
\usepackage{latexsym}
\usepackage[T1]{fontenc}
\usepackage[utf8]{inputenc}
\usepackage{microtype}
\usepackage{inconsolata}
\usepackage{graphicx}
\usepackage{amsmath}
\usepackage{booktabs}
\usepackage[normalem]{ulem}
\usepackage{amssymb}
\usepackage{dblfloatfix}
\usepackage{enumitem}
\usepackage{float}
\usepackage[section]{placeins}
\usepackage{multirow}
\usepackage[table]{xcolor}
\usepackage{subfigure}
\usepackage{svg}
\usepackage{comment}
\usepackage{soul}

\title{SG-LegalCite: A Principle-Augmented Benchmark for Legal Citation Retrieval in Singapore Law}

\author{
\normalfont
Shannon Lee Yueh Ern\textsuperscript{1},
Kaidong Feng\textsuperscript{1},
Yingpeng Du\textsuperscript{2},
Chloe Lee En Jia\textsuperscript{3},
\\
and Zhu Sun\textsuperscript{1}
\\[0.3em]
\textsuperscript{1}Singapore University of Technology and Design, Singapore
\\
\textsuperscript{2}Nanyang Technological University, Singapore
\\
\textsuperscript{3}Singapore Management University, Yong Pung How School of Law, Singapore
\\[0.2em]
\small{
\texttt{yuehernshannon\_lee@mymail.sutd.edu.sg},
\texttt{kaidong\_feng@sutd.edu.sg},
\texttt{yingpeng.du@ntu.edu.sg}
}
\\
\small{
\texttt{chloelee.2021@law.smu.edu.sg},
\texttt{zhu\_sun@sutd.edu.sg}
}
}

\begin{document}
\maketitle

\begin{abstract}
 
Legal citation in common-law systems depends not only on factual similarity, but also on the legal principle for which a precedent is invoked. However, existing benchmarks for legal citation retrieval use case facts, citation context or full judgments as inputs, where the governing legal principle is often missing or only implicitly expressed and entangled with broader context. As a result, models may retrieve precedents that are factually similar yet doctrinally irrelevant.  
This limitation is particularly consequential in Singapore, where the legal system has evolved independently: only domestic precedents are binding, while foreign authorities serve merely as persuasive references. %Accordingly, doctrinal alignment cannot be assumed even when factual patterns are similar.
%To address these gaps, 
Thus, we propose a new retrieval paradigm that ranks cited cases based on queries integrating case facts and explicit legal principle, inspired by the real-world legal reasoning workflows.
To support this paradigm, we further introduce SG-LegalCite, a dataset of 100,890 case–principle pairs extracted from 8,523 Singapore Supreme Court judgments spanning 2000–2025. Experiments across 11 baselines demonstrate the effectiveness of our principle-augmented retrieval paradigm, showing that explicit legal principles provide strong discriminative signals for legal citation retrieval. Our dataset and code is at \url{https://github.com/anonymousmeowmeow/SG-LegalCite}

\end{abstract}

\section{Introduction}
In common-law systems, legal citation is central to adjudication under \emph{stare decisis}: courts justify outcomes by applying, distinguishing, or refining principles established in prior decisions~\cite{woon1999precedent}. This has motivated growing interest in legal citation retrieval, which aims to identify precedents that support the reasoning of a citing judgment~\cite{farber2020citation,goebel2024coliee}. Typically, whether a precedent should be cited depends not only on factual similarity, but also on the legal principle for which it is invoked: cases with similar facts may implicate different doctrines and therefore support different lines of authority.

However, existing benchmarks for legal citation retrieval, as summarized in Table~\ref{tab:benchmarks}, always use case facts~\cite{li2024lecardv2,li2023muser,gao2024enhancing,santosh2024ecthrpcr}, citation context~\cite{hou2025clerc,wrzalik2021gerdalir,mahari2024lepard}, or full judgments~\cite{goebel2024coliee,mandal2017fire,joshi2023ucreat} as queries, where the governing legal principle is often missing, only implicitly expressed, or entangled with broader context. Consequently, such systems may retrieve precedents that are factually similar yet doctrinally irrelevant. This reveals a fundamental limitation: existing benchmarks fail to explicitly model the doctrinal basis of legal citation.

\begin{table*}[t]
  \centering
  \setlength{\tabcolsep}{2pt}
  \scriptsize
  %\vspace{-0.1in}
  \begin{tabular}{l|l|l|l|l|l}
    \toprule
    \textbf{Paradigm} & \textbf{Dataset} & \textbf{Query} & \textbf{Output} & \textbf{Jurisdiction} & \textbf{Reference} \\
    \midrule
    \multirow{6}{*}{Facts $\rightarrow$ Similar / Cited Cases}
      & AILA & Raw fact from judgment & Similar cases & India & \cite{bhattacharya2019aila} \\
      & LeCaRD & Raw fact from judgment & Similar cases & China & \cite{ma2021lecard} \\
      & MUSER & Raw fact + court opinion from judgment & Similar cases & China & \cite{li2023muser} \\
      & LeCaRDv2 & Raw fact from judgment & Similar cases & China & \cite{li2024lecardv2} \\
      & LEAD & Fact summary via LLMs from judgment & Similar cases & China & \cite{gao2024enhancing} \\
      & ECtHR-PCR & Raw fact field from judgment & Cited cases & Europe & \cite{santosh2024ecthrpcr} \\
    \hline
    \multirow{3}{*}{Citation Context $\rightarrow$ Cited Cases}
      & GerDaLIR & Citation paragraph without cited cases & Cited cases & Germany & \cite{wrzalik2021gerdalir} \\
      & LePaRD & Text before quotation (up to 300 words) & Quoted sentence & US & \cite{mahari2024lepard} \\
      & CLERC & 150 words before and after cited cases & Cited cases & US & \cite{hou2025clerc} \\\hline
    \multirow{3}{*}{Judgment $\rightarrow$ Cited Cases}
      & FIRE IRLeD & Full judgment (cited cases removed) & Cited cases & India & \cite{mandal2017fire} \\
      & IL-PCR & Full judgment (cited cases removed) & Cited cases & India & \cite{joshi2023ucreat} \\
      & COLIEE & Full judgment (cited cases removed) & Cited cases & Canada & \cite{goebel2024coliee} \\
    \hline
      \multirow{2}{*}{[Facts + Legal Principle] $\rightarrow$ Cited Cases}
       &\multirow{2}{*}{SG-LegalCite} & Fact summary via LLMs from judgment & \multirow{2}{*}{Cited cases} & \multirow{2}{*}{Singapore} &--\\
       & & + key legal principle & & \\ 
    \bottomrule
  \end{tabular}
  \vspace{-0.1in}
  \caption{Legal retrieval benchmarks comparison. 
  }
  \label{tab:benchmarks}
  \vspace{-0.15in}
\end{table*}

This limitation is particularly consequential in Singapore. Although Singapore follows the common-law tradition, its legal system has evolved independently: only domestic precedents are binding, while foreign authorities are merely persuasive. As a result, doctrinal alignment cannot be assumed even when factual patterns are similar. For instance, in negligence law, English courts apply the \emph{Caparo} framework~\cite{satyan2015case}, whereas Singapore adopts the two-stage approach in \emph{Spandeck}~\cite{spandeck2007}. A fact-based retrieval system may therefore return precedents that are factually relevant but doctrinally inapposite. Despite this, no existing benchmark covers legal citation retrieval in Singapore.

To fill the gap, we propose a new retrieval paradigm where cited cases are ranked based on queries that integrate case facts with explicit legal principles, reflecting real-world legal reasoning workflows.
To support such a paradigm, we introduce \textbf{SG-LegalCite}, a Singapore-specific dataset constructed from 8,523 Supreme Court judgments spanning 2000--2025. Through a multi-stage extraction pipeline, we obtain 100,890 case–principle pairs linking cited cases to the legal principles for which they are invoked. It enables controlled comparison between fact-only and principle-augmented retrieval, allowing direct evaluation of the role of legal principles.
We further evaluate SG-LegalCite across 11 baselines. Results consistently show that principle-augmented retrieval outperforms methods that do not explicitly model legal principles, indicating that such principles provide strong discriminative signals for legal citation retrieval.

In summary, we establish a principle-augmented benchmark for legal citation retrieval in Singapore law. Our main contributions are threefold: (1) \textbf{A new retrieval paradigm}: we formulate legal citation retrieval as a principle-aware retrieval task that integrates case facts with legal principles; (2) \textbf{A new legal dataset}: we introduce SG-LegalCite, the first large-scale benchmark for legal citation retrieval in Singapore; (3) \textbf{Comprehensive evaluation}: we benchmark diverse retrieval methods and show that incorporating explicit legal principles substantially improves retrieval performance.

\section{Related Work}

\subsection{Legal Retrieval Benchmarks}\label{sec:benchmarks}

Legal citation retrieval systems typically retrieve precedents that best support the legal reasoning of a citing judgment, given a query derived from that case. 
Existing legal retrieval benchmarks can be categorized into three groups based on the type of query they use, as summarized in Table~\ref{tab:benchmarks}. 

One line of research adopts case facts as queries to retrieve similar or cited cases. For instance, AILA~\cite{bhattacharya2019aila}, LeCaRD~\cite{ma2021lecard}, and LeCaRDv2~\cite{li2024lecardv2} use raw facts from judgments to retrieve similar cases. Later, MUSER~\cite{li2023muser} introduces a multi-view query formulation (i.e., raw facts plus court opinions), while LEAD~\cite{gao2024enhancing} leverages LLM-generated fact summaries from judgments instead of raw facts. Moreover, ECtHR-PCR~\cite{santosh2024ecthrpcr} adopts raw facts to retrieve cited cases and shows that \textit{fact-based queries are more effective than full judgments for retrieval}.
Another line of work exploits citation contexts (e.g., the text surrounding a cited case) as queries to retrieve cited cases, as in GerDaLIR~\cite{wrzalik2021gerdalir} and CLERC~\cite{hou2025clerc}, or to retrieve quoted sentences, as in LePaRD~\cite{mahari2024lepard}.
A third line of research directly uses the full judgment as the query to retrieve cited cases, as in COLIEE~\cite{goebel2024coliee}, FIRE-IRLeD~\cite{mandal2017fire}, and IL-PCR~\cite{joshi2023ucreat}.

Overall, existing benchmarks suffer from two key limitations. First, their queries--whether case facts, citation contexts, or full judgments--do not explicitly encode the legal principle for which a precedent is invoked. Second, they focus primarily on jurisdictions such as Canada, India, China, Europe, and US, with no coverage of Singapore. Although Singapore belongs to the common-law tradition, its legal system has evolved independently: only domestic precedents are binding, while foreign authorities serve merely as persuasive references.

\begin{figure*}[t]
\centering
\subfigure{
\includegraphics[width=0.95\columnwidth]{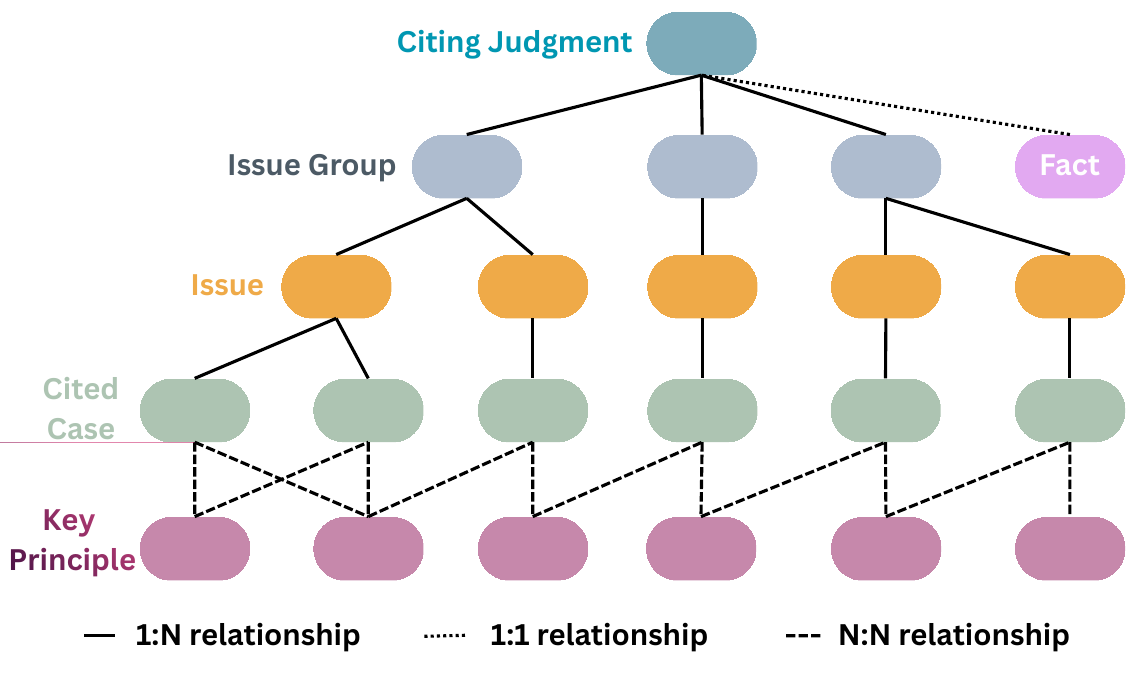}
\hspace{0.15in}
\subfigure{
\includegraphics[width=1\columnwidth]{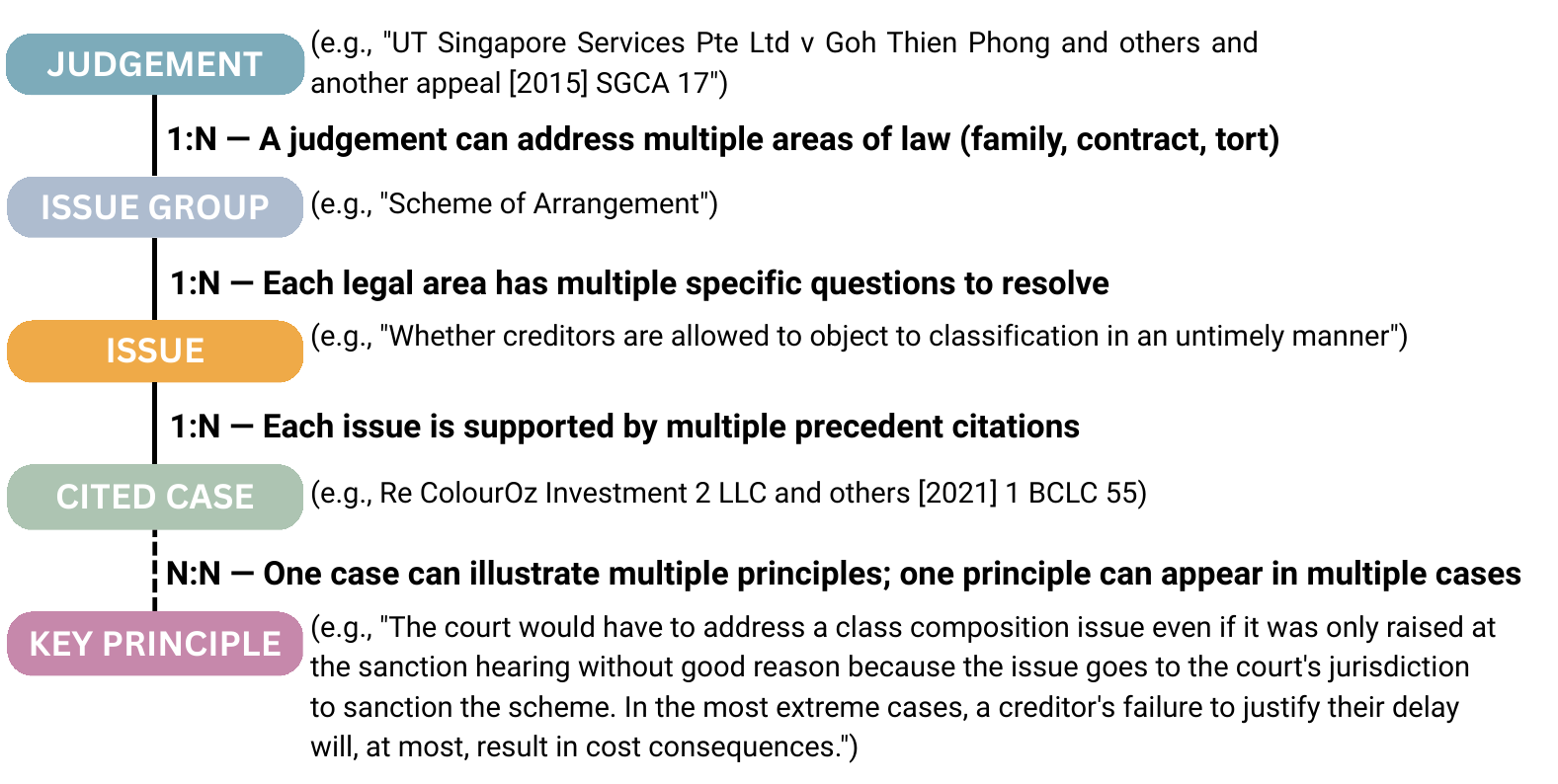}
}
}    
\vspace{-0.25in}
\caption{(Left) Data schema of SG-LegalCite; (Right) Example instances following the data schema.}
\label{fig:structure}
\vspace{-0.15in}
\end{figure*}

\subsection{Legal Citation Recommendation}

Citation recommendation aims to predict useful references from a query context. In the general domain, prior work includes context-aware probabilistic models~\cite{he2010context}, content-based neural methods~\cite{bhagavatula2018content,ebesu2017neural,jeong2020context}, and graph-based approaches that use citation structure \cite{cohan2020specter,farber2020citation}. However, in the legal domain, citation recommendation remains under-explored. Existing studies mostly predict legal citations from local textual context within judgments, including work on US veterans' appeals decisions~\cite{huang2021context}, prototype-based methods~\cite{luo2023prototype}, instruction-tuned models~\cite{wang2024empowering}, and recent graph-based approaches for German and Austrian legal data \cite{wendlinger2026missing}. 
However, these methods differ from our setting in that they predict citations either from local textual context within judgments (e.g., citation contexts) or from co-citation relationships among similar citing judgments, rather than retrieving cases using queries that explicitly integrate case facts with legal principles.

\section{The Proposed Retrieval Paradigm}
\label{sec:task}

Given a query $q$, we formulate legal citation retrieval as the task of retrieving relevant cited cases from a candidate pool $\mathcal{C} = \{c_1, c_2, \dots, c_N\}$. 
Each query consists of the factual description of the citing judgment together with the legal principle under which authority is sought, denoted $q = [\textit{facts};\; \textit{principle}]$. 
The model generates a ranked list over $\mathcal{C}$, with the goal of ranking the ground-truth cited cases $\mathcal{C}^{+} \subset \mathcal{C}$ as highly as possible.
Formally, the task is to learn a scoring function $\mathcal{F}(q,c)$ such that $\mathcal{F}(q,c^{+}) > 
\mathcal{F}(q,c^{-})$ for relevant cases $c^{+}\in\mathcal{C}^{+}$ and irrelevant cases $c^{-}\in\mathcal{C}\setminus\mathcal{C}^{+}$.
Our formulation differs from prior legal retrieval settings in that the query explicitly includes the legal principle associated with the citation, enabling retrieval to be conditioned not only on case facts but also on the doctrinal basis for which authority is invoked.

\section{Dataset Construction}\label{sec:dataset}

\subsection{Data Schema}
SG-LegalCite is constructed to support our proposed retrieval paradigm, %in Section~\ref{sec:task}, 
where the query combines case facts with explicit legal principles. 
Accordingly, each record in SG-LegalCite contains five fields: the factual summary of the citing judgment ($f$), the legal principle for which a precedent is invoked ($k$), the cited case ($c$), together with a %\sout{high-level doctrinal category} 
\textcolor{black}{fine-grained doctrinal tag} 
(\textit{issue group}; e.g., ``Damages'', ``Contract'') and a specific legal issue (\textit{issue}; e.g., ``Whether the duty of care was breached'').
Specifically, the triplet $(f,k,c)$ directly instantiates our paradigm in Section~\ref{sec:task}: $f$ and $k$ form the query, and $c$ is the retrieval target.  
The issue provides the legal question being resolved in the judgment, providing necessary context to identify the correct principle within potentially long citation passages and avoid  peripheral statements. In addition, issue and issue group serve as metadata: issue group enables filtering and evaluation by %\sout{legal domain (e.g., Damages)} 
\textcolor{black}{doctrinal tag (e.g., Damages)}, while issue supports finer-grained analysis of model performance and error patterns across specific doctrinal questions.

Figure~\ref{fig:structure} illustrates the data schema, showing hierarchical relationships among judgments, fact, issue groups, issues, cited cases, and key principles. Each judgment contributes one shared facts field $f$ and refers to multiple individual citations each associated with its own principle $k$, cited case $c$, issue and issue group. This structure reflects the process of legal reasoning: the facts describe the dispute as a whole, whereas individual citations are invoked for legal principles relevant to particular aspects of that dispute.

\begin{figure*}[t]
  \centering
  % \includesvg[width=0.9\textwidth]{image/legalpipelineV4.svg}
  \includegraphics[width=0.99\textwidth]{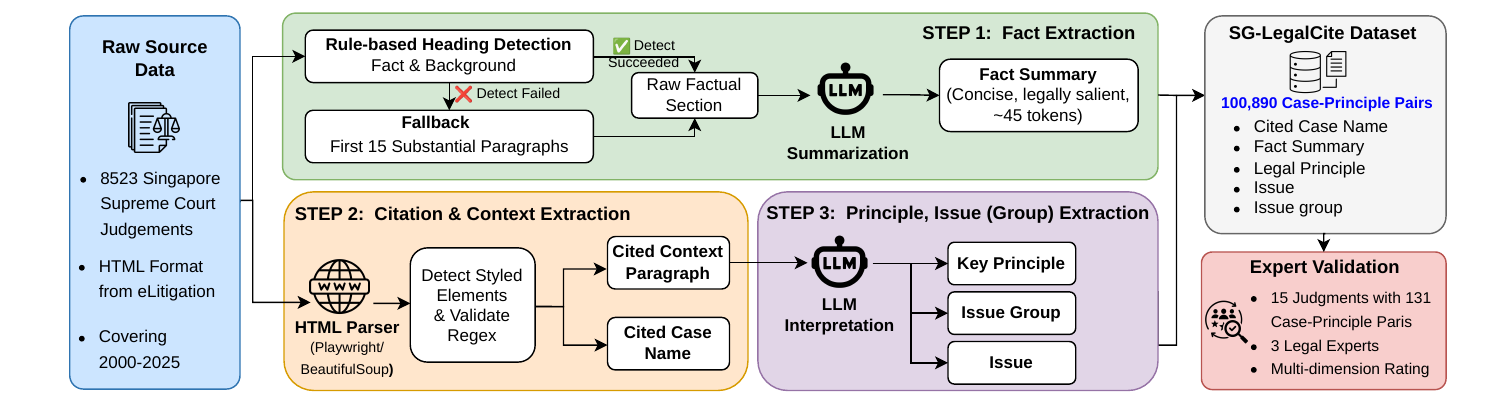}
  \vspace{-0.1in}
  \caption{SG-LegalCite dataset construction pipeline.}
  \label{fig:pipeline}
  \vspace{-0.15in}
\end{figure*}

\subsection{Data Source}
We collected 8,523 Supreme Court judgments spanning 2000--2025 from the eLitigation platform\footnote{https://www.elitigation.sg}, the official public repository for Singapore court judgments. The corpus covers the Court of Appeal, Court of Appeal International, and High Court (including Registrar and Family divisions). \textcolor{black}{Specifically, eLitigation's full archive contains 10,372 judgments (2000-2005) and we identified 9,404 valid judgment URLs. Of these, 8,523 judgments contained at least one cited case and were retained; 881 judgments contained zero cited cases and were dropped as they cannot contribute case--principle pairs to the retrieval task. 
%The retained judgments yield the 100,890 case--principle pairs that constitute SG-LegalCite.
}
%
%\textcolor{red}{For context, eLitigation's full archive contains 10,627 judgments (from 2000 through partial 2026, as of 6 May 2026). To ensure comprehensive coverage of our target window, we performed exhaustive URL enumeration, systematically probing every case number from 1 upward for each combination of year (2000--2025) and Supreme Court type (SGHC, SGCA, SGHCF, SGHCR, SGCAI), stopping after 10 consecutive missing case numbers per (year, court) block. This identified 9,404 valid judgment URLs, representing an effectively exhaustive enumeration of publicly accessible Singapore Supreme Court judgments on eLitigation for the 2000--2025 period. Of these, 8,523 judgments contained at least one cited case and were retained; 881 judgments contained zero cited cases and were dropped as they cannot contribute case--principle pairs to the retrieval task. The retained judgments yield the 100,890 case--principle pairs that constitute SG-LegalCite.}
%
Singapore judgments follow a broadly consistent structure: introduction and background sections describe the parties, procedural posture, and facts, while later issue-specific analysis sections contain the court's reasoning and citations to precedent. This regularity makes it feasible to separate factual background from citation-bearing doctrinal discussion, which is essential for constructing triplets of the form $(f,k,c)$.

\subsection{Extraction Pipeline}
Since the corpus has 8,523 judgments, full manual annotation would require substantial expert effort (estimated at approximately \textbf{4-5 hours} per citing judgment) and is therefore impractical at this scale. We thus propose an automatic construction pipeline using large language model (LLM)~\cite{liu2024deepseek} driven extraction with few-shot prompting~\cite{mao2025reinforced}, followed by expert validation on a sampled subset to assess quality. Figure~\ref{fig:pipeline} shows the pipeline, consisting of three main steps.

\smallskip\noindent\textbf{Step 1: Fact Extraction.}
For each judgment, we first locate the factual section through rule-based heading detection over the HTML structure. We prioritise headings such as \textit{Facts}, \textit{Background}, \textit{Introduction}, and \textit{Dispute}, and fall back to alternative factual headings where necessary. If no reliable heading is found, we extract the first 15 substantial paragraphs from the opening part of the judgment. Duplicated text is removed before summarization.

Raw factual sections are often lengthy and interspersed with procedural history and rhetorical detail that are irrelevant to the retrieval task.
So, we truncate the text to 3{,}000 words and then summarize it with LLMs (e.g., Deepseek) via a structured prompt\footnote{All the prompts utilized in the data construction process are provided in Appendix~\ref{sec:fact_prompt} and Appendix~\ref{sec:prompt}.}. The LLM is asked to produce a concise 2--3 sentence, lawyer-style summary that preserves legally salient facts while removing procedural and rhetorical detail. This step reduces the average fact length from 1,034 tokens to 45 tokens, yielding compact factual descriptions suitable for retrieval. 
\textcolor{black}{This compression level is designed to reflect how legal practitioners formulate retrieval queries in practice, that is, a concise and legally salient summary rather than the full factual background; it is also necessary to fit input limits of several encoder baselines (e.g., Legal-BERT, SBERT)}.

% \textcolor{red}{This compression level is designed to reflect how legal practitioners formulate retrieval queries in practice — a concise, legally salient summary rather than the full factual background — and is also necessary to fit within the 512-token input limits of several encoder baselines (e.g., Legal-BERT, SBERT), since the fact summary is concatenated with additional fields (legal principle, issue, and issue group) in our principle-augmented and ablation settings.}

\smallskip\noindent\textbf{Step 2: Citation and Context Extraction.}
We use Playwright\footnote{https://playwright.dev} for web scraping and BeautifulSoup\footnote{https://beautiful-soup-4.readthedocs.io} for HTML parsing. Cited case names are identified through styled HTML elements (e.g., \texttt{<em>}, \texttt{<i>}), with citation formats validated against Singapore neutral citation patterns (e.g., [2024] SGHC 123) and common foreign patterns. For each detected citation, we retain the surrounding $\pm 5$ paragraphs as its local context.

\smallskip\noindent\textbf{Step 3: Principle, Issue and Issue Group Extraction.}
Legal principles are rarely stated as self-contained sentences. Instead, they are typically embedded in the court's reasoning alongside case comparisons, factual distinctions, and policy discussion. Extracting them thus requires interpreting the argumentative structure of the citation context, which is difficult to capture with rule-based or pattern-matching methods.
We thus use an LLM to extract three 
fields from each citation paragraph extracted from step 2: (1) \textbf{Key Principles Illustrated} -- the 
legal principle in full formal phrasing as stated in the judgment; 
(2) \textbf{Issue Group} -- \textcolor{black}{fine-grained doctrinal tag} (e.g., ``Damages'', ``Contract'');
and (3) \textbf{Issue} -- precise legal question (e.g., ``Whether the duty of care 
was breached''). 
The LLM is instructed to preserve explicit legal formulations when present in the source text, with prompt examples spanning diverse legal 
domains (e.g., matrimonial assets, insolvency and statutory interpretation) to 
ensure robust extraction across case types.
The extracted legal principle is then used as the principle field $k$ in our retrieval task.

\begin{table}[t]
  \scriptsize
  \setlength{\tabcolsep}{2pt}
  \centering
  \begin{tabular}{l|cc}
    \toprule
    Model & HSS & Cost per Case \\
    \midrule
    Claude Sonnet 4~\cite{anthropic2025claude4} & \textbf{91.3\%} & \$0.24 \\
    DeepSeek-V3~\cite{liu2024deepseek} & 86.7\% & \textbf{\$0.02} \\
    GPT-4o~\cite{openai2024gpt4o} & 84.7\% & \$0.16 \\
    \bottomrule
  \end{tabular}
  \vspace{-0.1in}
  \caption{Performance comparison of different LLMs on case-principle pair extraction task.}
  \label{tab:llm_selection}
  \vspace{-0.1in}
\end{table}

\begin{table}[t]
  \scriptsize
  \centering
  \setlength{\tabcolsep}{8pt}
  % \vspace{-0.1in}
  \begin{tabular}{l|cc}
    \toprule
    \# Examples & HSS (Issue) & HSS (Key Principle) \\
    \midrule
    0-shot & 74.7\% & 86.7\% \\
    5-shot & 76.7\% & 84.7\% \\
    10-shot & 79.3\% & 85.3\% \\
    15-shot & \textbf{80.7\%} & \textbf{86.7}\% \\
    20-shot & 79.3\% & 86.7\% \\
    \bottomrule
  \end{tabular}
  \vspace{-0.1in}
  \caption{Performance comparison across different numbers of in-context examples for issue and key-principle extraction using DeepSeek-V3.}
  \label{tab:fewshot}
  \vspace{-0.15in}
\end{table}

\subsection{LLM Selection}\label{sec:llm_sel}

To select the extraction model, we compare three LLMs (i.e., Claude Sonnet 4, DeepSeek-V3, and GPT-4o) against 25 judgments containing 725 case-principle pairs annotated by legal experts. 
We evaluate model outputs against these annotations using a \textit{Hybrid Similarity Score (HSS)}, defined as the equally weighted average of ROUGE~\cite{lin2004rouge}, which captures lexical overlap in legal terminology, and BERTScore~\cite{zhang2020bertscore}, which captures semantic similarity. Through this evaluation, we aim to determine which LLM provides the best accuracy–cost trade-off and how many in-context examples should be included in the prompt. 

We first compare all three LLMs under the same prompt setting. In Table~\ref{tab:llm_selection}, Claude achieves the highest accuracy (91.3\%) at \$0.24 per case, whereas DeepSeek attains 86.7\% at \$0.02 per case -- a 12$\times$ cost reduction for only a 4.6\% drop in HSS. DeepSeek thus offers the most favourable accuracy–cost trade-off among the three LLMs.
We then further tune the number of in-context examples on DeepSeek. Table~\ref{tab:fewshot} shows that key principle extraction already performs strongly in the zero-shot setting (86.7\%), with little variation across shot counts, consistent with legal principles often being stated relatively explicitly in judicial reasoning. Issue extraction, by contrast, benefits more from additional examples, improving from 74.7\% in the zero-shot setting to 80.7\% at 15-shot. So, we  select DeepSeek with 15 examples for full-scale extraction, yielding a total cost of \$78.22.

\begin{table}[!t]
\centering
  \scriptsize
  \begin{tabular}{ll}
    \toprule
    Attribute & Value \\
    \midrule
    Time Span & 2000--2025 \\
    Unique Judgments & 8,523 \\
    Case-Principle Pairs & 100,890 \\
    Unique Principles & 72,500 \\
    Unique Cited Cases & 48,478 \\
    Unique Issues & 86,519 \\
    Unique Issue Groups & 9,748 \\
    Avg.\ Raw Fact Length & 1,034.4 tokens \\
    Avg.\ Fact Length (post-summary) & 45.1 tokens \\
    Avg.\ Citation Paragraph Length & 1,100.5 tokens \\
    Avg.\ Principle Length & 69.9 tokens \\
    \bottomrule
  \end{tabular}
  \vspace{-0.1in}
  \caption{SG-LegalCite dataset statistics.}
  \label{tab:dataset_stats}
  \vspace{-0.15in}
\end{table}

\subsection{The Dataset and Expert Validation}\label{sec:expert_validation}

\smallskip\noindent\textbf{Constructed Dataset.}
Table~\ref{tab:dataset_stats} shows the statistics of the resulting dataset. 
\textcolor{black}{To characterise the doctrinal composition of the corpus, we classified all 8,523 judgments against the 34 practice-area tags used by Singapore Law Watch~\cite{singaporelawwatch}. The most represented primary domains are Criminal Law (20.7\%), Business \& Commerce (10.9\%), Civil Law \& Procedure (8.5\%), Tort (6.8\%), and Family Law (6.8\%); the full distribution is provided in Appendix~\ref{app:domain-distribution}. We did not artificially balance the corpus across domains. SG-LegalCite is intended as a benchmark for retrieval systems deployed by Singapore legal practitioners. The natural distribution, dominated by criminal and commercial matters, with sparse coverage of emerging areas such as data protection and health care \& life sciences, is itself the population a deployed system would face. The observed skew reflects the institutional role of the Supreme Court (apex criminal jurisdiction; commercial-hub case mix) and is consistent with how the Singapore Law Reports themselves catalogue judgments.}
%
% \textcolor{red}{To characterise the doctrinal composition of the corpus, we classified all 8,523 judgments against the 34 practice-area tags used by Singapore Law Watch~\cite{singaporelawwatch}. The most represented primary domains are Criminal Law (20.7\%), Business \& Commerce (10.9\%), Civil Law \& Procedure (8.5\%), Tort (6.8\%), and Family Law (6.8\%); the full distribution is provided in Appendix~\ref{app:domain-distribution}. We did not artificially balance the corpus across domains. SG-LegalCite is intended as a benchmark for retrieval systems deployed by Singapore legal practitioners, and the natural distribution of Supreme Court cases — dominated by criminal and commercial matters, with sparse coverage of emerging areas such as data protection and health care \& life sciences — is itself the population a deployed system would face. The observed skew reflects the institutional role of the Supreme Court (apex criminal jurisdiction; commercial-hub case mix) and is consistent with how the Singapore Law Reports themselves catalogue judgments.} 
%
Moreover, Table~\ref{tab:extraction_example} in Appendix~\ref{app:example-record} provides a concrete example of a citation-level record produced by our pipeline, illustrating how raw factual and citation context are transformed into the structured fields in SG-LegalCite.

\smallskip\noindent\textbf{Quality Assessment.}
Expert validation is conducted on a randomly sampled subset to assess the extraction quality. Specifically, three legal experts (details in Appendix~\ref{app:data-construction}) independently rated 131 case-principle pairs from 15 randomly sampled judgments using a 5-point scale across various dimensions on the key principle illustrated, issue and issue group (details in Appendix~\ref{app:data-construction}). 
The overall mean ratings are 3.22 ($\sigma=1.10$) for Key Principles Illustrated, 3.93 ($\sigma=0.98$) for Issue, and 4.28 ($\sigma=0.99$) for Issue Group, with $ICC(2,k)=0.54$, indicating moderate inter-rater agreement. We regard this as consistent with the interpretive nature of legal principle formulation: a single citation may admit multiple defensible doctrinal formulations even when annotators broadly agree on the underlying legal point.

\smallskip\noindent\textbf{Comparison Across LLMs.}
To further validate the LLM selection in Section~\ref{sec:llm_sel}, the same three annotators independently rated 131 case--principle pairs extracted by Claude Sonnet 4 from the same 15 judgments under identical 1--5 scoring criteria, without being informed which model produced each set of extractions. Results show Claude achieving higher quality scores than DeepSeek: Key Principles Illustrated improved from 3.22 to 3.65 (+13.4\%), Issue extraction from 3.93 to 4.35 (+10.7\%), and Issue Group from 4.28 to 4.49 (+4.9\%). Despite the quality improvements, DeepSeek-V3's substantially lower API cost makes it more practical for large-scale extraction (see Table~\ref{tab:llm_selection}). We therefore retain DeepSeek-V3 as the extraction model for SG-LegalCite.

\section{Experiments}\label{sec:experiments}

We conduct experiments on our SG-LegalCite using 11 state-of-the-art methods to evaluate the effectiveness of our proposed retrieval paradigm.

\begin{table*}[t]
  \centering
  \scriptsize
  \setlength{\tabcolsep}{5pt}
  \begin{tabular}{lccccc|ccccc}
    \toprule
    & \multicolumn{5}{c}{Fact-only Query} & \multicolumn{5}{c}{Principle-augmented Query} \\
    \cmidrule(lr){2-6} \cmidrule(lr){7-11}
    Model & MRR & R@1 & R@5 & R@10 & R@20 & MRR & R@1 & R@5 & R@10 & R@20 \\
    \midrule
    \multicolumn{11}{l}{\textit{\colorbox{gray!15}{Conventional Lexical Baseline}}} \\
    BM25               &  1.8 &  0.6 &  2.1 &  3.2 &  4.9 &  3.2\textsuperscript{\,+79\%}  &  2.0\textsuperscript{\,+221\%} &  3.5\textsuperscript{\,+65\%}  &  4.7\textsuperscript{\,+45\%}  &  6.6\textsuperscript{\,+35\%} \\
    \midrule
    \multicolumn{11}{l}{\textit{\colorbox{gray!15}{Conventional Pre-trained Language Models}}} \\
    SBERT              & 10.5 &  4.7 & 15.1 & 22.5 & 31.2 & 20.9\textsuperscript{\,+99\%}  & 12.9\textsuperscript{\,+174\%} & 27.8\textsuperscript{\,+84\%}  & 36.3\textsuperscript{\,+61\%}  & 45.8\textsuperscript{\,+47\%} \\
    Legal-BERT         &  6.2 &  2.2 &  8.0 & 13.5 & 21.1 & 14.1\textsuperscript{\,+128\%} &  7.1\textsuperscript{\,+228\%} & 19.6\textsuperscript{\,+145\%} & 27.5\textsuperscript{\,+104\%} & 37.5\textsuperscript{\,+78\%} \\
    Custom Legal-BERT  &  6.1 &  2.1 &  8.0 & 13.7 & 20.8 &  9.2\textsuperscript{\,+51\%}  &  4.0\textsuperscript{\,+90\%}  & 12.4\textsuperscript{\,+55\%}  & 18.9\textsuperscript{\,+38\%}  & 27.4\textsuperscript{\,+32\%} \\
    Legal-Longformer   &  6.8 &  2.6 &  9.0 & 14.4 & 22.2 & 10.0\textsuperscript{\,+47\%}  &  4.4\textsuperscript{\,+69\%}  & 13.9\textsuperscript{\,+54\%}  & 20.9\textsuperscript{\,+45\%}  & 29.9\textsuperscript{\,+35\%} \\
    Pile-of-Law BERT   &  5.7 &  2.0 &  7.3 & 12.2 & 18.9 & 12.4\textsuperscript{\,+118\%} &  5.8\textsuperscript{\,+190\%} & 17.5\textsuperscript{\,+140\%} & 25.5\textsuperscript{\,+109\%} & 35.2\textsuperscript{\,+86\%} \\
    SAILER   &  0.8 &  0.1 &  0.5 &  1.0 &  2.1 &  0.9\textsuperscript{\,+13\%}  &  0.2\textsuperscript{\,+100\%} &  0.6\textsuperscript{\,+20\%}  &  1.2\textsuperscript{\,+20\%}  &  2.0\textsuperscript{\,$-$5\%} \\
    Legal-en-RoBERTa   &  6.9 &  2.6 &  9.1 & 14.6 & 21.9 &  9.4\textsuperscript{\,+36\%}  &  3.8\textsuperscript{\,+46\%}  & 13.0\textsuperscript{\,+43\%}  & 20.0\textsuperscript{\,+37\%}  & 29.6\textsuperscript{\,+35\%} \\
    \midrule
    \multicolumn{11}{l}{\textit{\colorbox{gray!15}{Large-scale Legal Language Models}}} \\
    AdaptLLM           & 11.6 &  5.3 & 16.4 & 24.7 & 33.8 & 29.5\textsuperscript{\,+154\%} & 17.9\textsuperscript{\,+238\%} & 41.6\textsuperscript{\,+154\%} & 53.5\textsuperscript{\,+117\%} & 64.9\textsuperscript{\,+92\%} \\
    Lawma-8B           & 10.0 &  4.0 & 14.1 & 21.7 & 31.7 & 30.5\textsuperscript{\,+205\%} & 18.2\textsuperscript{\,+355\%} & 43.8\textsuperscript{\,+211\%} & 56.6\textsuperscript{\,+161\%} & 68.6\textsuperscript{\,+116\%} \\
    SaulLM-7B          & \textbf{13.0} &  \textbf{5.9} & \textbf{18.7} & \textbf{27.3} & \textbf{37.5} & \textbf{38.2}\textsuperscript{\,\textbf{+194\%}} & \textbf{24.4}\textsuperscript{\,\textbf{+314\%}} & \textbf{54.2}\textsuperscript{\,\textbf{+190\%}} & \textbf{66.7}\textsuperscript{\,\textbf{+144\%}} & \textbf{77.2}\textsuperscript{\,\textbf{+106\%}} \\
    \bottomrule
  \end{tabular}
  \vspace{-0.1in}
  \caption{The Performance (\%) of 11 baselines on our SG-LegalCite Dataset. Relative gains (shown as percentages in the upper-right corner of the MRR or Recall values for the principle-augmented query) are computed against the corresponding fact-only setting for the same model. The best results for each paradigm are highlighted in bold.}
  \label{tab:results_ft}
  \vspace{-0.1in}
\end{table*}

\subsection{Experiment Setup}

\noindent\textbf{Baselines.} 
We compare 11 methods in three categories: \textit{conventional lexical models},
\textit{conventional pre-trained language models} and \textit{large-scale legal language models}. The \textbf{first category} includes one classic retrieval model BM25~\cite{rosa2021yes}. The \textbf{second category} includes one general-purpose baseline -- SBERT~\cite{reimers2019sbert}, and six legal-specific models: Legal-BERT~\cite{chalkidis2020legal}, Custom Legal-BERT~\cite{zheng2021does}, Pile-of-Law BERT\cite{henderson2022pile}, SAILER~\cite{li2023sailer}, Legal-Longformer~\cite{chalkidis2023lexfiles}, and Legal-EN-RoBERTa~\cite{chalkidis2023lexfiles}. The \textbf{third category} has three models tailored for the legal domain: AdaptLLM~\cite{cheng2024adaptllm}, SaulLM-7B~\cite{colombo2024saullm}, and Lawma-8B~\cite{dominguez2024lawma}.  This diverse set of models spans variations in parameter scale, pre-training corpus size, and pre-training objectives, thus enabling a comprehensive evaluation\footnote{We did not compare with Lawyer GPT~\cite{yao2024lawyergpt} and InternLM-Law~\cite{fei2025internlmlaw} due to unavailability of released models; and did not compare with SaulLM-54B~\cite{colombo2024saullm} due to GPU memory limits.}.

\smallskip\noindent\textbf{Evaluation Protocols.} Following SOTA retrieval methods~\cite{hou2025clerc}, we use Recall@$k$ (R@$k$) and Mean Reciprocal Rank (MRR) as evaluation metrics, capturing both citation retrieval accuracy and ranking quality. $k$ is set to \{1, 5, 10, 20\}.
Results are reported from a single run per model due to the high computational cost of fine-tuning 11 models across different query settings. To ensure reproducibility, we fix the random seed to 42.

\smallskip\noindent\textbf{Query Settings.}
We fine-tune the 11 selected approaches on our  SG-LegalCite dataset under two query settings: \textbf{\textit{fact-only}} and \textbf{\textit{principle-augmented}}. In the fact-only setting, the retriever receives only the factual background of the citing judgment as the query. In the principle-augmented setting, the query concatenates the judgment facts with the corresponding legal principle, exposing both the factual pattern and the doctrinal basis of the citation. To speed up the testing process, we constructed a candidate pool: all models are tested on a 1000-way candidate pool consisting of one gold cited case and 999 randomly sampled negatives.

\smallskip\noindent\textbf{Implementation Details.} For fine-tuning, we use contrastive learning with symmetric InfoNCE loss~\cite{oord2018representation} over query–case pairs. The dataset is split into train, validation, and test sets with the 8:1:1 ratio. 
All models are trained with batch size 64 and learning 
rate 2e-5. 
The temperature parameter in InfoNCE loss is initialized at 0.07 and learned during training, and optimal hyperparameter settings are based on validation loss. All our experiments are conducted on 
NVIDIA A100 GPUs with 80 GB memory.

\subsection{Main Results}\label{sec:main-result}

Table~\ref{tab:results_ft} reports the performance of all models fine-tuned on our SG-LegalCite under the two query settings: \textit{fact-only} and \textit{principle-augmented}. Several major findings are noted. 

(1) \textbf{\textit{Large-scale legal language models consistently outperform smaller models across both query settings}}. SaulLM-7B achieves the best performance, followed by SBERT, whereas BM25 performs the worst. This suggests legal citation retrieval benefits substantially from both model scale and domain-specific pre-training, enabling better modeling of relationships among case facts, legal principles, and precedents.

(2) \textbf{\textit{Principle-augmented retrieval consistently outperforms fact-only retrieval}}. Across nearly all models, augmenting queries with legal principles yields substantial gains. On average, it improves MRR by 111\% and Recall by 124\%, suggesting that explicit legal principles provide additional semantics that facilitate accurate cited-case retrieval.

(3) \textbf{\textit{Large-scale legal language models benefit most from principle augmentation}}. On average, our paradigm improves MRR by 79\%, 70\%, and 184\%, and Recall by 111\%, 77\%, and 183\% across the three model categories. The widened performance gap suggests stronger models better exploit principle-level semantics beyond lexical or sentence-level matching.

(4) \textcolor{black}{\textbf{\textit{Among conventional pre-trained language models, the general-purpose SBERT unexpectedly outperforms all legal-specific models}}. It may be attributed to SBERT’s retrieval-aligned contrastive pretraining, while most legal encoders rely on masked language modeling. Besides, differences in corpus language, scale, recency, and jurisdictional coverage may limit the transferability of legal-specific models to Singapore law. Further studies are needed to isolate the contribution of each factor and we leave them as our future work.} 

(5) \textbf{Recall@$k$ improvements show a consistent trend across models}. Gains are largest at Recall@1 and gradually decrease as $k$ increases (184\% $\rightarrow$ 106\% $\rightarrow$ 80\% $\rightarrow$ 60\% for $k=1,5,10,20$ on average). The strongest improvements at top ranks indicate that principle-augmented queries are particularly effective at ranking the correct cited case first.

\begin{table}[h]
  \centering
  \scriptsize
  \setlength{\tabcolsep}{1pt}
  \textcolor{black}{
  \begin{tabular}{lccccc}
    \toprule
    \textbf{Query Setting} & \textbf{MRR} & \textbf{R@1} & \textbf{R@5} & \textbf{R@10} & \textbf{R@20} \\
    \midrule
    \multicolumn{6}{l}{\textit{\colorbox{gray!15}{SBERT (all-mpnet-base-v2)}}} \\
    Fact-only             & 10.5 & 4.7 & 15.1 & 22.5 & 31.2 \\
    Fact + Issue Group    & 19.4\textsuperscript{\,+85\%}   & 10.7\textsuperscript{\,+127\%}  & 27.5\textsuperscript{\,+82\%}  & 37.4\textsuperscript{\,+66\%}  & 49.0\textsuperscript{\,+57\%} \\
    Fact + Issue          & 20.8\textsuperscript{\,+98\%}   & 11.5\textsuperscript{\,+145\%}  & 29.6\textsuperscript{\,+96\%}  & 39.7\textsuperscript{\,+76\%}  & 50.2\textsuperscript{\,+61\%} \\
    Fact + Principle      & 20.9\textsuperscript{\,+99\%}   & 12.9\textsuperscript{\,+174\%}  & 27.8\textsuperscript{\,+84\%}  & 36.3\textsuperscript{\,+61\%}  & 45.8\textsuperscript{\,+47\%} \\
    Fact + All      & \textbf{26.3}\textsuperscript{\,\textbf{+151\%}} & \textbf{16.5}\textsuperscript{\,\textbf{+251\%}} & \textbf{36.6}\textsuperscript{\,\textbf{+142\%}} & \textbf{46.1}\textsuperscript{\,\textbf{+105\%}} & \textbf{55.8}\textsuperscript{\,\textbf{+79\%}} \\
    Fact + Context   & 18.1\textsuperscript{\,+72\%}  &  9.4\textsuperscript{\,+100\%} & 26.3\textsuperscript{\,+74\%}  & 35.8\textsuperscript{\,+59\%}  & 46.7\textsuperscript{\,+50\%} \\
    \midrule
    \multicolumn{6}{l}{\textit{\colorbox{gray!15}{SaulLM-7B (Equall/Saul-Instruct-v1)}}} \\
    Fact-only             & 13.0 & 5.9 & 18.7 & 27.3 & 37.5 \\
    Fact + Issue Group    & 17.5\textsuperscript{\,+35\%}   & 8.8\textsuperscript{\,+50\%}    & 25.3\textsuperscript{\,+35\%}  & 35.5\textsuperscript{\,+30\%}  & 47.1\textsuperscript{\,+26\%} \\
    Fact + Issue          & 36.3\textsuperscript{\,+179\%}  & 22.1\textsuperscript{\,+275\%}  & 52.8\textsuperscript{\,+182\%} & 65.9\textsuperscript{\,+141\%} & 77.1\textsuperscript{\,+106\%} \\
    Fact + Principle      & \textbf{38.2}\textsuperscript{\,\textbf{+194\%}} & \textbf{24.4}\textsuperscript{\,\textbf{+314\%}} & \textbf{54.2}\textsuperscript{\,\textbf{+190\%}} & \textbf{66.7}\textsuperscript{\,\textbf{+144\%}} & \textbf{77.2}\textsuperscript{\,\textbf{+106\%}} \\
    Fact + All      & 34.8\textsuperscript{\,+168\%}  & 21.9\textsuperscript{\,+271\%}  & 49.3\textsuperscript{\,+164\%} & 61.5\textsuperscript{\,+125\%} & 72.9\textsuperscript{\,+94\%} \\
    Fact + Context   & 18.9\textsuperscript{\,+45\%}  &  9.6\textsuperscript{\,+63\%}  & 27.4\textsuperscript{\,+47\%}  & 38.1\textsuperscript{\,+40\%}  & 49.9\textsuperscript{\,+33\%} \\
    \bottomrule
  \end{tabular}
  }
  \vspace{-0.1in}
  \caption{\textcolor{black}{Detailed Ablation Study. Relative gains (shown as superscript percentages) are computed against the corresponding fact-only setting. %for the same model. 
  %All values are percentages (\%). The best results per model are highlighted in bold.
  %Field granularity ablation. All five settings evaluate on identical candidate pools (1 gold + 999 random negatives, 9{,}979 pools per setting) with identical hyperparameters; only the query construction differs. Relative gains (shown as superscript percentages) are computed against the corresponding fact-only setting for the same model. All values are percentages (\%). The best results per model are highlighted in bold.}
  }}
  \label{tab:granularity-ablation}
  \vspace{-0.2in}
\end{table}

\subsection{Detailed Ablation Study}
% \textcolor{red}{A further question raised by our main results is whether the gains from principle-augmented retrieval reflect the specific value of formal doctrinal phrasing, or whether any structured legal field --- including coarser ones such as the issue group --- would provide comparable signal. To answer this, we conduct a controlled ablation across the full range of doctrinal granularities present in SG-LegalCite, comparing four single-field augmentations against a multi-field combination.}
\textcolor{black}{\textbf{Principle vs. Other Doctrinal Signals}.
Our main results in Section~\ref{sec:main-result} establish that augmenting queries with the legal principle consistently improves retrieval performance. A natural follow-up question is whether this gain reflects the particular discriminative value of formally phrased legal principles, or whether any structured doctrinal signal extracted from the same citation context would yield comparable improvements. So, we compare four types of doctrinal augmentation derived from SG-LegalCite against the fact-only baseline:
(i) Fact + Issue Group, augmenting with the fine-grained doctrinal tag; (ii) Fact + Issue, augmenting with the specific legal question; (iv) Fact + Principle, augmenting with the extracted legal principle; and (v) Fact + All, concatenating all the three doctrinal signals. Detailed experimental settings are in Appendix~\ref{app-granularity-ablation}. We evaluate them on two representative models: SBERT and SaulLM-7B.
}

% \smallskip\noindent\textbf{Setup.} \textcolor{red}{We evaluate five query settings on the same candidate pools (1 gold cited case + 999 random negatives, 9{,}979 pools): (i) \emph{fact-only}; (ii) \emph{fact + issue group}, augmenting with the coarsest doctrinal tag (e.g., ``Damages''); (iii) \emph{fact + issue}, augmenting with the specific legal question (e.g., ``Whether the duty of care was breached''); (iv) \emph{fact + principle}, augmenting with the formal legal principle as cited; and (v) \emph{fact + all fields}, concatenating issue group, issue, and principle. To ensure strict apples-to-apples comparison, all five conditions share identical test pools deduplicated on (fact, principle, cited case), identical negative samples, identical hyperparameters (batch size 64, learning rate 2e-5, 10 epochs with early stopping, QLoRA $r{=}16$ for SaulLM-7B), and only the query construction differs. We evaluate on two representative models spanning the conventional encoder family (SBERT) and large-scale legal LLM family (SaulLM-7B).}

\textcolor{black}{Four observations emerge from Table~\ref{tab:granularity-ablation}. First, \textit{\textbf{every form of doctrinal augmentation improves over the fact-only baseline across both models}}, validating the broader value of structured legal signals at any granularity. 
Second, \textit{\textbf{the benefit of granularity is mediated by model capacity}}: SaulLM-7B is highly sensitive to signal specificity (Issue Group: +35\% MRR; Issue: +179\%; Principle: +194\%), while SBERT gains comparably from all three single-signal settings (+85–99\% MRR), suggesting that capable models exploit fine-grained doctrinal precision that weaker encoders cannot
Third, \textit{\textbf{"Fact + Principle" is the strongest single doctrinal signal for SaulLM-7B}}, confirming that formally phrased legal principles carry discriminative information compared with other signals. 
Finally, \textit{\textbf{the "Fact + All" configuration is best for SBERT but underperforms "Fact + Principle" for SaulLM-7B}}, indicating that concatenating overlapping signals provides complementary cues for weaker encoders but dilutes the already-sufficient principle signal for stronger models.}

\smallskip\noindent\textcolor{black}{
\textbf{Principle vs. Citation Context}. To further verify the efficacy of our proposed paradigm, we compare the extracted principles against raw citation context at matched text budgets. Details on citation context extraction is in Appendix~\ref{app:settings-citation-context}.
We evaluate two representative models (SBERT and SaulLM-7B) under the "Fact + Context" setting.
Table~\ref{tab:granularity-ablation} (last row of each model) shows the results. 
"Fact + Context" improves substantially over "Fact-only" retrieval (SBERT: +72\% MRR; SaulLM-7B: +45\% MRR), confirming that citation-proximal text carries useful retrieval signal. However, "Fact + Principle" consistently outperform "Fact + Context" at the same text budget: SBERT achieves +15\% higher MRR and +37\% higher R@1 with principles, while SaulLM-7B shows +102\% higher MRR and +154\% higher R@1. The advantage is especially pronounced for SaulLM-7B, suggesting that larger models are better able to exploit the distilled doctrinal content in the principle field. These results demonstrate that the principle extraction step captures genuine doctrinal signal beyond what raw citation context provides, and that the gains reported in Table~\ref{tab:results_ft} are not attributable solely to richer input text.}

\smallskip\noindent
\textcolor{black}{
\textbf{Gold Principle vs. Predicted Principle}.
Our main experiments use the gold legal principle extracted from the citing judgment, which represents an oracle setting where the doctrinal basis of the citation is known a priori. In reality, a lawyer may begin with only a factual description of the case and may not have identified the governing principle. To assess whether our principle-augmented paradigm is robust to this cold-start scenario, we test the fine-tuned SaulLM-7B under our "Fact + Principle" paradigm across three query settings: (i) Fact-only using fact as the query; (ii) Cold-start using fact and the predicted principle; and (iii) Gold (oracle) using fact and the extracted principle in our constructed datasets.  
Specifically, we randomly sample 50 cases from the test set. For each, we use DeepSeek-V3 to predict the legal principle from the fact summary alone\footnote{\textcolor{black}{The principle prediction prompt is in Appendix~\ref{app:coldstart-prompt}.}}, with case names disallowed in the output to prevent leakage. During inference, only the query text differs while all other settings are kept the same to ensure a fair comparison.
}

% \noindent\textbf{Setup.} \textcolor{red}{We randomly sample 50 pools from the test set (seed=42). For each, we use DeepSeek-V3\footnote{\textcolor{red}{The cold-start principle prediction prompt is provided in Appendix~\ref{app:coldstart-prompt}.}} to predict the legal principle from the fact summary alone, with case names disallowed in the output to prevent leakage. The same fine-tuned SaulLM-7B is then evaluated on the same 50 candidate pools (1 gold case + 999 random negatives) under three conditions: (i) \emph{fact-only}, using \texttt{[FACT]~\textit{fact}} as the query; (ii) \emph{cold-start}, using \texttt{[FACT]~\textit{fact}~[PRINCIPLE]~\textit{predicted principle}}; and (iii) \emph{gold} (oracle), using \texttt{[FACT]~\textit{fact}~[PRINCIPLE]~\textit{gold principle}}. Only the query text differs across the three conditions; the model, candidate pools, and ground-truth cases are held constant.}

\textcolor{black}{As shown in Table~\ref{tab:coldstart}, "Cold-start" substantially outperforms "Fact-only" (+28\% MRR, +40\% R@1), confirming that the principle-augmented paradigm provides real benefit even when principles must be predicted from facts alone. The gap between "Cold-start" and "Gold"  indicates clear headroom for improvement in principle prediction, e.g., through fine-tuned principle predictors or lawyer-in-the-loop refinement.}

\begin{table}[t]
\centering
\scriptsize
\setlength{\tabcolsep}{3pt}
\textcolor{black}{
\begin{tabular}{lccccc}
\toprule
\textbf{Setting} & \textbf{MRR} & \textbf{R@1} & \textbf{R@5} & \textbf{R@10} & \textbf{R@20} \\
\midrule
Fact-only          & 17.1                                & 10.0                                & 24.0                                & 28.0                                & 38.0 \\
Cold-start         & 21.9\textsuperscript{\,+28\%}       & 14.0\textsuperscript{\,+40\%}       & 28.0\textsuperscript{\,+17\%}       & 34.0\textsuperscript{\,+21\%}       & 44.0\textsuperscript{\,+16\%} \\
Gold (oracle)      & \textbf{49.5}\textsuperscript{\,\textbf{+190\%}} & \textbf{36.0}\textsuperscript{\,\textbf{+260\%}} & \textbf{66.0}\textsuperscript{\,\textbf{+175\%}} & \textbf{74.0}\textsuperscript{\,\textbf{+164\%}} & \textbf{78.0}\textsuperscript{\,\textbf{+105\%}} \\
\bottomrule
\end{tabular}
}
\vspace{-0.1in}
\caption{\textcolor{black}{Gold principle vs. predicted principle. 
%Cold-start evaluation of fine-tuned SaulLM-7B on 50 randomly sampled test pools. Relative gains (shown as percentages in the upper-right corner) are computed against the fact-only setting. All values are percentages (\%). Same 50 pools, same model — only the query text differs.
}}
\label{tab:coldstart}
\vspace{-0.2in}
\end{table}

\section{Conclusion and Future Work}

We introduce SG-LegalCite, a principle-augmented benchmark for legal citation retrieval. First, we propose a novel retrieval paradigm that integrates legal principles with case facts, moving beyond fact-only methods. Second, to support such a paradigm, we construct the first large-scale dataset for Singapore law, comprising 100,890 case–principle pairs from 8,523 Supreme Court judgments (2000–2025). Experiments show the effectiveness of our proposed paradigm, demonstrate that explicit legal principles provide strong discriminative signals for citation retrieval and underscore the importance of modeling both factual similarity and doctrinal relevance. 

\textbf{Limitations}.
%This paper has several limitations. 
First, it relies on LLM-generated summaries and extracted principles with only partial expert validation, which may introduce biases and propagate upstream errors into retrieval performance. Second, experiments are conducted on a sampled candidate pool (1,000 cases) \textcolor{black}{for inference efficiency, which may not fully reflect large-scale retrieval settings. Future work will explore more effective candidate selection methods and evaluation on the full candidate pool}. Lastly, results are reported from a single run per model with a fixed random seed due to computational constraints, limiting the assessment of variance and robustness.
% This paper has several limitations. First, it relies on LLM-generated summaries and extracted principles with partial expert validation, introducing potential biases and propagation of upstream errors into retrieval performance. \textcolor{red}{Second, the experiments are conducted on a sampled candidate pool (1,000 cases) to improve inference efficiency, which may not fully reflect real-world large-scale retrieval settings. In future work, we plan to develop more effective candidate selection methods or evaluate it using the entire case pool}
% %
% Lastly, results are reported from a single run per model (with a fixed random seed) due to computational constraints, limiting the assessment of variance and robustness.

% \smallskip\noindent\textbf{Future Work}. 
% Building on SG-LegalCite, several directions remain. First, future work can explore generation-based citation recommendation, where LLMs generate candidate citations conditioned on facts and legal principles, with retrieval used for grounding to ensure faithfulness. Second, richer use of metadata such as issue and issue group could enable \sout{domain-aware} \textcolor{red}{doctrine-aware} ranking, hierarchical retrieval, and multi-task learning. Finally, developing interpretable and trustworthy systems, particularly those providing principle-grounded explanations aligned with specific issues, will be essential for real-world legal applications.

\section*{Acknowledgment}
LLMs were used to support coding and manuscript polishing. All scientific design, methodology, and analysis are the authors’ own work.

%\section*{Ethical Considerations}

\bibliography{references}

@article{mao2025reinforced,
  title={Reinforced prompt personalization for recommendation with large language models},
  author={Mao, Wenyu and Wu, Jiancan and Chen, Weijian and Gao, Chongming and Wang, Xiang and He, Xiangnan},
  journal={ACM Transactions on Information Systems},
  volume={43},
  number={3},
  pages={1--27},
  year={2025},
  publisher={ACM New York, NY}
}

@article{liu2024deepseek,
  title={Deepseek-v3 technical report},
  author={Liu, Aixin and Feng, Bei and Xue, Bing and Wang, Bingxuan and Wu, Bochao and Lu, Chengda and Zhao, Chenggang and Deng, Chengqi and Zhang, Chenyu and Ruan, Chong and others},
  journal={arXiv preprint arXiv:2412.19437},
  year={2024}
}

@misc{openai2024gpt4o,
  author       = {{OpenAI}},
  title        = {Introducing GPT-4o},
  year         = {2024},
  howpublished = {\url{https://openai.com/index/hello-gpt-4o/}}
}

@misc{anthropic2025claude4,
  author       = {{Anthropic}},
  title        = {Introducing Claude 4},
  year         = {2025},
  howpublished = {\url{https://www.anthropic.com/news/claude-4}}
}

@article{rosa2021yes,
  title={Yes, bm25 is a strong baseline for legal case retrieval},
  author={Rosa, Guilherme Moraes and Rodrigues, Ruan Chaves and Lotufo, Roberto and Nogueira, Rodrigo},
  journal={arXiv preprint arXiv:2105.05686},
  year={2021}
}

@incollection{woon1999precedent,
  author    = {Woon, Walter},
  title     = {The Doctrine of Judicial Precedent},
  booktitle = {The Singapore Legal System},
  editor    = {Tan, Kevin Y.L.},
  edition   = {2nd},
  year      = {1999},
  publisher = {Singapore University Press},
  address   = {Singapore},
}

@inproceedings{chalkidis2020legal,
  author    = {Chalkidis, Ilias and Fergadiotis, Manos and Malakasiotis, Prodromos and Aletras, Nikolaos and Androutsopoulos, Ion},
  title     = {{LEGAL-BERT}: The Muppets straight out of Law School},
  booktitle = {Findings of the Association for Computational Linguistics: EMNLP 2020},
  pages     = {2898--2904},
  year      = {2020},
  publisher = {Association for Computational Linguistics},
  address   = {Online},
}

@inproceedings{chalkidis2023lexfiles,
  author    = {Chalkidis, Ilias and Garneau, Nicolas and Goanta, Catalina and Katz, Daniel and S{\o}gaard, Anders},
  title     = {{LeXFiles} and {LegalLAMA}: Facilitating {E}nglish Multinational Legal Language Model Development},
  booktitle = {Proceedings of the 61st Annual Meeting of the Association for Computational Linguistics},
  pages     = {15513--15535},
  year      = {2023},
  address   = {Toronto, Canada},
  publisher = {Association for Computational Linguistics},
}

@inproceedings{zheng2021does,
  author    = {Zheng, Lucia and Guha, Neel and Anderson, Brandon R. and Henderson, Peter and Ho, Daniel E.},
  title     = {When Does Pretraining Help? {A}ssessing Self-Supervised Learning for Law and the {CaseHOLD} Dataset of 53,000+ Legal Holdings},
  booktitle = {Proceedings of the 18th International Conference on Artificial Intelligence and Law},
  pages     = {159--168},
  year      = {2021},
  publisher = {ACM},
  address   = {New York, NY, USA},
}

@misc{spandeck2007,
  author    = {{Singapore Court of Appeal}},
  title     = {Spandeck Engineering (S) Pte Ltd v Defence Science \& Technology Agency},
  year      = {2007},
  note      = {[2007] SGCA 37},
}

@article{satyan2015case,
  title={Case analysis: Caparo Industries Plc v. Dickman},
  author={Satyan, Kanika},
  journal={Dickman (July 5, 2015)},
  year={2015}
}

@article{goebel2024coliee,
  author  = {Goebel, Randy and Kano, Yoshinobu and Kim, Mi-Young and Rabelo, Juliano and Satoh, Ken and Yoshioka, Masaharu},
  title   = {Overview and Discussion of the Competition on Legal Information Extraction/Entailment ({COLIEE}) 2023},
  journal = {The Review of Socionetwork Strategies},
  volume  = {18},
  number  = {1},
  pages   = {27--47},
  year    = {2024},
  doi     = {10.1007/s12626-023-00152-0},
}

@inproceedings{henderson2022pile,
  title     = {Pile of Law: Learning Responsible Data Filtering from the Law and a 256GB Open-Source Legal Dataset},
  author    = {Henderson, Peter and Krass, Mark S and Zheng, Lucia and Guha, Neel and Manning, Christopher D and Jurafsky, Dan and Ho, Daniel E},
  booktitle = {Advances in Neural Information Processing Systems},
  volume    = {35},
  year      = {2022},
  publisher = {Curran Associates, Inc.},
  address   = {Red Hook, NY, USA},
}

@inproceedings{li2023sailer,
  title     = {{SAILER}: Structure-aware Pre-trained Language Model for Legal Case Retrieval},
  author    = {Li, Haitao and Ai, Qingyao and Chen, Jia and Dong, Qian and Wu, Yueyue and Liu, Yiqun and Chen, Chong and Tian, Qi},
  booktitle = {Proceedings of the 46th International ACM SIGIR Conference on Research and Development in Information Retrieval},
  pages     = {1035--1044},
  year      = {2023},
  publisher = {ACM},
  address   = {New York, NY, USA},
}

@inproceedings{colombo2024saullm,
  title     = {{SaulLM-54B} \& {SaulLM-141B}: Scaling Up Domain Adaptation for the Legal Domain},
  author    = {Colombo, Pierre and Pires, Telmo Pessoa and Boudiaf, Malik and Melo, Rui and Culver, Dominic and Morgado, Sofia and Malaboeuf, Etienne and Hautreux, Gabriel and Charpentier, Johanne and Desa, Michael},
  booktitle = {Advances in Neural Information Processing Systems},
  year      = {2024},
  publisher = {Curran Associates, Inc.},
  address   = {Red Hook, NY, USA},
}

@inproceedings{joshi2023ucreat,
  title     = {{U-CREAT}: Unsupervised Case Retrieval using Events ex{tA}cTion},
  author    = {Joshi, Abhinav and Sharma, Akshat and Tanikella, Sai Kiran and Modi, Ashutosh},
  booktitle = {Proceedings of the 61st Annual Meeting of the Association for Computational Linguistics (Volume 1: Long Papers)},
  month     = jul,
  year      = {2023},
  address   = {Toronto, Canada},
  publisher = {Association for Computational Linguistics},
  url       = {https://aclanthology.org/2023.acl-long.777/},
  doi       = {10.18653/v1/2023.acl-long.777},
}

@inproceedings{li2023muser,
  title     = {{MUSER}: A Multi-View Similar Case Retrieval Dataset},
  author    = {Li, Qingquan and Hu, Yiran and Yao, Feng and Xiao, Chaojun and Liu, Zhiyuan and Sun, Maosong and Shen, Weixing},
  booktitle = {Proceedings of the 32nd ACM International Conference on Information and Knowledge Management},
  year      = {2023},
  pages     = {5336--5340},
  publisher = {ACM},
  doi       = {10.1145/3583780.3615125},
}

@inproceedings{gao2024enhancing,
  title     = {Enhancing Legal Case Retrieval via Scaling High-quality Synthetic Query-Candidate Pairs},
  author    = {Gao, Cheng and Xiao, Chaojun and Liu, Zhenghao and Chen, Huimin and Liu, Zhiyuan and Sun, Maosong},
  booktitle = {Proceedings of the 2024 Conference on Empirical Methods in Natural Language Processing},
  month     = nov,
  year      = {2024},
  address   = {Miami, Florida, USA},
  publisher = {Association for Computational Linguistics},
  url       = {https://aclanthology.org/2024.emnlp-main.402/},
  doi       = {10.18653/v1/2024.emnlp-main.402},
  pages     = {7086--7100},
}

@inproceedings{wrzalik2021gerdalir,
  title     = {{G}er{D}a{LIR}: A {G}erman Dataset for Legal Information Retrieval},
  author    = {Wrzalik, Marco and Krechel, Dirk},
  booktitle = {Proceedings of the Natural Legal Language Processing Workshop 2021},
  month     = nov,
  year      = {2021},
  address   = {Punta Cana, Dominican Republic},
  publisher = {Association for Computational Linguistics},
  url       = {https://aclanthology.org/2021.nllp-1.13/},
  doi       = {10.18653/v1/2021.nllp-1.13},
  pages     = {123--128},
}

@inproceedings{ma2021lecard,
  title     = {{LeCaRD}: A Legal Case Retrieval Dataset for {C}hinese Law System},
  author    = {Ma, Yixiao and Shao, Yunqiu and Wu, Yueyue and Liu, Yiqun and Zhang, Ruizhe and Zhang, Min and Ma, Shaoping},
  booktitle = {Proceedings of the 44th International ACM SIGIR Conference on Research and Development in Information Retrieval},
  year      = {2021},
  pages     = {2342--2348},
  publisher = {ACM},
  doi       = {10.1145/3404835.3463250},
}

@inproceedings{li2024lecardv2,
  title     = {{LeCaRDv2}: A Large-Scale {C}hinese Legal Case Retrieval Dataset},
  author    = {Li, Haitao and Shao, Yunqiu and Wu, Yueyue and Ai, Qingyao and Ma, Yixiao and Liu, Yiqun},
  booktitle = {Proceedings of the 47th International ACM SIGIR Conference on Research and Development in Information Retrieval},
  year      = {2024},
  pages     = {2405--2409},
  publisher = {ACM},
  doi       = {10.1145/3626772.3657887},
}

@inproceedings{mandal2017fire,
  title     = {Overview of the {FIRE} 2017 {IRLeD} Track: Information Retrieval from Legal Documents},
  author    = {Mandal, Arpan and Ghosh, Kripabandhu and Bhattacharya, Arnab and Pal, Arindam and Ghosh, Saptarshi},
  booktitle = {FIRE (Working Notes)},
  year      = {2017},
  pages     = {63--68},
}

@inproceedings{bhattacharya2019aila,
  title     = {Overview of the {FIRE} 2019 {AILA} Track: Artificial Intelligence for Legal Assistance},
  author    = {Bhattacharya, Paheli and Ghosh, Kripabandhu and Ghosh, Saptarshi and Pal, Arindam and Mehta, Parth and Bhattacharya, Arnab and Majumder, Prasenjit},
  booktitle = {FIRE (Working Notes)},
  year      = {2019},
  pages     = {1--12},
}

@inproceedings{santosh2024ecthrpcr,
  title     = {{ECtHR-PCR}: A Dataset for Precedent Understanding and Prior Case Retrieval in the {E}uropean Court of Human Rights},
  author    = {T.y.s.s., Santosh and Haddad, Rashid and Grabmair, Matthias},
  booktitle = {Proceedings of the 2024 Joint International Conference on Computational Linguistics, Language Resources and Evaluation (LREC-COLING)},
  month     = may,
  year      = {2024},
  address   = {Torino, Italia},
  publisher = {ELRA and ICCL},
  url       = {https://aclanthology.org/2024.lrec-main.486/},
  pages     = {5473--5483},
}

@inproceedings{hou2025clerc,
  title     = {{CLERC}: A Dataset for {U.S.} Legal Case Retrieval and Retrieval-Augmented Analysis Generation},
  author    = {Hou, Abe Bohan and Weller, Orion and Qin, Guanghui and Yang, Eugene and Lawrie, Dawn and Holzenberger, Nils and Blair-Stanek, Andrew and Van Durme, Benjamin},
  booktitle = {Findings of the Association for Computational Linguistics: NAACL 2025},
  year      = {2025},
  pages     = {7898--7913},
  address   = {Albuquerque, New Mexico},
  publisher = {Association for Computational Linguistics},
  url       = {https://aclanthology.org/2025.findings-naacl.441/},
}

@inproceedings{mahari2024lepard,
  title     = {{LePaRD}: A Large-Scale Dataset of Judicial Citations to Precedent},
  author    = {Mahari, Robert and Stammbach, Dominik and Ash, Elliott and Pentland, Alex},
  booktitle = {Proceedings of the 62nd Annual Meeting of the Association for Computational Linguistics (Volume 1: Long Papers)},
  year      = {2024},
  month     = aug,
  pages     = {9863--9877},
  address   = {Bangkok, Thailand},
  publisher = {Association for Computational Linguistics},
  url       = {https://aclanthology.org/2024.acl-long.532/},
  doi       = {10.18653/v1/2024.acl-long.532},
}

@inproceedings{lin2004rouge,
  title={{ROUGE}: A Package for Automatic Evaluation of Summaries},
  author={Lin, Chin-Yew},
  booktitle={Text Summarization Branches Out},
  pages={74--81},
  year={2004},
  address={Barcelona, Spain},
  publisher={Association for Computational Linguistics},
  url={https://aclanthology.org/W04-1013/}
}

@inproceedings{zhang2020bertscore,
  title={{BERTScore}: Evaluating Text Generation with {BERT}},
  author={Zhang, Tianyi and Kishore, Varsha and Wu, Felix and Weinberger, Kilian Q. and Artzi, Yoav},
  booktitle={International Conference on Learning Representations},
  year={2020},
  url={https://openreview.net/forum?id=SkeHuCVFDr}
}

@inproceedings{reimers2019sbert,
  title     = {Sentence-{BERT}: Sentence Embeddings using Siamese {BERT}-Networks},
  author    = {Reimers, Nils and Gurevych, Iryna},
  booktitle = {Proceedings of the 2019 Conference on Empirical Methods in Natural Language Processing and the 9th International Joint Conference on Natural Language Processing},
  pages     = {3982--3992},
  year      = {2019},
  month     = nov,
  address   = {Hong Kong, China},
  publisher = {Association for Computational Linguistics},
  url       = {https://aclanthology.org/D19-1410/},
  doi       = {10.18653/v1/D19-1410},
}

@inproceedings{cheng2024adaptllm,
  title={Adapting Large Language Models via Reading Comprehension},
  author={Cheng, Daixuan and Huang, Shaohan and Wei, Furu},
  booktitle={International Conference on Learning Representations},
  year={2024}
}

@inproceedings{yao2024lawyergpt,
  title={Lawyer {GPT}: A Legal Large Language Model with Enhanced Domain Knowledge and Reasoning Capabilities},
  author={Yao, Shunyu and Ke, Qingqing and Wang, Qiwei and Li, Kangtong and Hu, Jie},
  booktitle={Proceedings of the 2024 3rd International Symposium on Robotics, Artificial Intelligence and Information Engineering},
  pages={108--112},
  year={2024},
  organization={ACM},
  doi={10.1145/3689299.3689319}
}

@inproceedings{fei2025internlmlaw,
  title={Intern{LM}-Law: An Open-Sourced {C}hinese Legal Large Language Model},
  author={Fei, Zhiwei and Zhang, Songyang and Shen, Xiaoyu and Zhu, Dawei and Wang, Xiao and Ge, Jidong and Ng, Vincent},
  booktitle={Proceedings of the 31st International Conference on Computational Linguistics},
  pages={9376--9392},
  year={2025},
  organization={Association for Computational Linguistics}
}

@misc{dominguez2024lawma,
  title     = {Lawma: The Power of Specialization for Legal Tasks},
  author    = {Dominguez-Olmedo, Ricardo and Nanda, Vedant and Abebe, Rediet and Bechtold, Stefan and Engel, Christoph and Frankenreiter, Jens and Gummadi, Krishna and Hardt, Moritz and Livermore, Michael},
  year      = {2024},
  eprint    = {2407.16615},
  archivePrefix = {arXiv},
  primaryClass  = {cs.CL},
  url       = {https://arxiv.org/abs/2407.16615},
}

@article{oord2018representation,
  title={Representation Learning with Contrastive Predictive Coding},
  author={van den Oord, A{\"a}ron and Li, Yazhe and Vinyals, Oriol},
  journal={arXiv preprint arXiv:1807.03748},
  year={2018}
}

@inproceedings{he2010context,
  author    = {He, Qi and Pei, Jian and Kifer, Daniel and Mitra, Prasenjit and Giles, C. Lee},
  title     = {Context-Aware Citation Recommendation},
  booktitle = {Proceedings of the 19th International Conference on World Wide Web},
  pages     = {421--430},
  year      = {2010},
  publisher = {ACM},
  address   = {New York, NY, USA}
}

@inproceedings{bhagavatula2018content,
  author    = {Bhagavatula, Chandra and Feldman, Sergey and Power, Russell and Ammar, Waleed},
  title     = {Content-Based Citation Recommendation},
  booktitle = {Proceedings of the 2018 Conference of the North American Chapter of the Association for Computational Linguistics: Human Language Technologies},
  pages     = {238--251},
  year      = {2018},
  publisher = {Association for Computational Linguistics},
}

@inproceedings{ebesu2017neural,
  author    = {Ebesu, Travis and Fang, Yi},
  title     = {Neural Citation Network for Context-Aware Citation Recommendation},
  booktitle = {Proceedings of the 40th International {ACM} {SIGIR} Conference on Research and Development in Information Retrieval},
  pages     = {1093--1096},
  year      = {2017},
  publisher = {ACM},
  address   = {New York, NY, USA},
  doi       = {10.1145/3077136.3080730},
}

@article{jeong2020context,
  author    = {Jeong, Chanwoo and Jang, Sion and Park, Eunjeong and Choi, Sungchul},
  title     = {A Context-Aware Citation Recommendation Model with {BERT} and Graph Convolutional Networks},
  journal   = {Scientometrics},
  volume    = {124},
  number    = {3},
  pages     = {1907--1922},
  year      = {2020},
  doi       = {10.1007/s11192-020-03561-y},
}

@inproceedings{cohan2020specter,
  author    = {Cohan, Arman and Feldman, Sergey and Beltagy, Iz and Downey, Doug and Weld, Daniel},
  title     = {{SPECTER}: Document-level Representation Learning using Citation-informed Transformers},
  booktitle = {Proceedings of the 58th Annual Meeting of the Association for Computational Linguistics },
  pages     = {2270--2282},
  year      = {2020},
  publisher = {Association for Computational Linguistics},
  doi       = {10.18653/v1/2020.acl-main.207},
}

@article{farber2020citation,
  author    = {F{\"a}rber, Michael and Jatowt, Adam},
  title     = {Citation Recommendation: Approaches and Datasets},
  journal   = {International Journal on Digital Libraries},
  volume    = {21},
  number    = {4},
  pages     = {375--405},
  year      = {2020},
  doi       = {10.1007/s00799-020-00288-2},
}

@inproceedings{huang2021context,
  author    = {Huang, Zihan and Low, Charles and Teng, Mengqiu and Zhang, Hongyi
               and Ho, Daniel E. and Krass, Mark S. and Grabmair, Matthias},
  title     = {Context-Aware Legal Citation Recommendation using Deep Learning},
  booktitle = {Proceedings of the 18th International Conference on Artificial
               Intelligence and Law},
  pages     = {79--88},
  year      = {2021},
  publisher = {ACM},
  address   = {New York, NY, USA},
  doi       = {10.1145/3462757.3466066},
}

@inproceedings{luo2023prototype,
  author    = {Luo, Chu Fei and Bhambhoria, Rohan and Dahan, Samuel and Zhu, Xiaodan},
  title     = {Prototype-Based Interpretability for Legal Citation Prediction},
  booktitle = {Findings of the Association for Computational Linguistics: {ACL} 2023},
  pages     = {4883--4898},
  year      = {2023},
  publisher = {Association for Computational Linguistics},
}

@inproceedings{wang2024empowering,
  author    = {Wang, Jie and Bansal, Kanha and Arapakis, Ioannis and Ge, Xuri
               and Jose, Joemon M.},
  title     = {Empowering Legal Citation Recommendation via Efficient
               Instruction-Tuning of Pre-trained Language Models},
  booktitle = {Proceedings of the 46th European Conference on Information
               Retrieval},
  pages     = {310--324},
  year      = {2024},
  publisher = {Springer},
  doi       = {10.1007/978-3-031-56027-9_19},
}

@inproceedings{wendlinger2026missing,
  author    = {Wendlinger, Lorenz and Nonn, Simon Alexander and Al Zubaer, Abdullah
               and Granitzer, Michael},
  title     = {The Missing Link: Joint Legal Citation Prediction using
               Heterogeneous Graph Enrichment},
  booktitle = {Proceedings of the 36th International Conference on Database and
               Expert Systems Applications},
  pages     = {197--211},
  year      = {2025},
  publisher = {Springer},
  doi       = {10.1007/978-3-032-02088-8_14},
}

@misc{singaporelawwatch,
  author       = {{Singapore Academy of Law}},
  title        = {Singapore Law Watch},
  howpublished = {\url{https://www.singaporelawwatch.sg/}},
  year         = {2026}
}

\appendix
%\clearpage

\section{Fact Extraction Prompt and Examples}
\label{sec:fact_prompt}

The following prompt was used for judgment-level fact extraction (Section~\ref{sec:dataset}). The model was instructed to produce a concise lawyer-style summary of the case facts, suitable for use as a retrieval query.

\smallskip
\noindent\textbf{Task instruction:} Rewrite the key facts of this case as a lawyer would describe their client's situation to a colleague --- conversational, practical, and focused on what matters legally.

\smallskip
\noindent\textbf{Rules:}
\begin{itemize}[leftmargin=1.5em,itemsep=0pt]
  \item Write 2--3 sentences maximum.
  \item Use phrases like ``my client'', ``the other party'', ``the company'' instead of full names.
  \item Focus on: what happened and what is at stake.
  \item Do not include legal questions or ask anything.
  \item Write as if you are a lawyer explaining the situation, not summarising a judgment.
  \item Do not start with any preamble such as ``Of course'', ``Sure'', or ``Based on the provided text''.
  \item Start directly with the facts.
  \item If the text does not contain enough information, respond with: ``Insufficient information to summarize facts.''
\end{itemize}

\smallskip
\noindent\textbf{In-context examples:}
\begin{itemize}[leftmargin=1.5em,itemsep=0pt]
  \item My client signed a 2-year clause preventing them from working for any competitor after leaving the company.
  \item My client transferred money by mistake to the wrong company and there is no contractual relationship between them.
  \item My client is going through a divorce after 15 years together. Her husband earned more but she was a stay-at-home mum taking care of three kids.
  \item A new bubble tea shop opened down the road with a name and cup design that customers keep mixing up with my client's brand.
  \item My client's customers have been accidentally buying a competitor's product thinking it was ours.
  \item The shareholders want to sue a director for losses from a failed project.
  \item We have medical reports showing our client's mental state was significantly impaired at the time of the offence.
  \item The document production exercise will cost our client over \$200,000 and it seems disproportionate to what is actually at stake.
\end{itemize}

\section{Principle, Issue and Issue Group Extraction Prompt and Examples}
\label{sec:prompt}

The following prompt was used for citation-level principle extraction (Section~\ref{sec:dataset}). The model was instructed to act as an expert legal analyst and extract three structured fields from each citation paragraph: \textit{Key Principles Illustrated}, \textit{Issue Group}, and \textit{Issue}. Output was required in TSV format for direct downstream processing.

\smallskip
\noindent\textbf{System prompt:} \textit{You are an expert legal analyst. Your job is to extract structured outputs for legal analytics.}

\smallskip
\noindent\textbf{Task instruction:} Given a paragraph from a judgment, extract exactly three fields: (1)~\textit{Key Principles} --- use the full formal legal phrasing reflected in the paragraph; do not simplify unless removing procedural noise is strictly necessary; (2)~\textit{Issue Group} --- the high-level legal category (e.g., ``Division of matrimonial assets'', ``Damages'', ``Scheme of Arrangement''); (3)~\textit{Issue} --- the precise legal issue in judgment style (prefer ``Whether\ldots'', ``When\ldots'', ``Identifying\ldots'', or ``Determining\ldots'').

\smallskip
\noindent\textbf{Output format:} A single TSV table with a header row and columns in the following order: \texttt{Cited Case}, \texttt{Paragraph}, \texttt{Key Principles Illustrated}, \texttt{Issue}, \texttt{Issue Group}. No additional keys, comments, or explanations outside the TSV. Full legal phrasing must be preserved; party names and citations are omitted unless essential to state the legal issue.

\smallskip
\noindent The 15 in-context examples used in the prompt are reproduced below.

\smallskip
\noindent\textbf{Example 1}

\noindent\textit{Input:} The Husband argues that the court should draw an adverse inference against the Wife for failing to disclose bank statements of her account with DBS Bank Ltd. Adverse inferences are appropriate where there is a substratum of evidence that establishes a prima facie case against the person whom the inference is to be drawn, and that person must have had some particular access to the information he or she is said to be hiding: \textit{UZN v UZM} [2021] 1 SLR 426 at [60]. I draw the adverse inference against the Wife that she had savings in the DBS Bank account. Adopting the uplift approach, I adjust the ratio for the division of the matrimonial assets by 5\% in favour of the Husband.

\noindent\textit{Output:}
\begin{itemize}[leftmargin=1.5em,itemsep=0pt]
  \item \textbf{Key Principles:} Adverse inferences are appropriate where there is a substratum of evidence that establishes a prima facie case against the person whom the inference is to be drawn, and that person must have had some particular access to the information he or she is said to be hiding.
  \item \textbf{Issue Group:} Division of matrimonial assets
  \item \textbf{Issue:} When adverse inferences should be drawn
\end{itemize}

\smallskip
\noindent\textbf{Example 2}

\noindent\textit{Input:} \textit{ColourOz} was decided after the release of the 2020 Practice Statement. There the English High Court reviewed the process for effecting a scheme of arrangement. Snowden J stated that: Whilst the court would always have to address a class question even if raised at sanction (because it goes to jurisdiction), the implicit warning now repeated in para 10 of the New Practice Statement is that unless a good reason can be shown, such a late submission is unlikely to be well received and might, in an extreme case, justify disallowing an opposing creditor's costs. In our view, the court would have to address a class composition issue even if it was only raised at the sanction hearing without good reason because the issue goes to the court's jurisdiction to sanction the scheme. Even in extreme cases, an objecting creditor's failure to justify their delay will, at most, result in cost consequences.

\noindent\textit{Output:}
\begin{itemize}[leftmargin=1.5em,itemsep=0pt]
  \item \textbf{Key Principles:} The court would have to address a class composition issue even if it was only raised at the sanction hearing without good reason because the issue goes to the court's jurisdiction to sanction the scheme. In the most extreme cases, a creditor's failure to justify their delay will, at most, result in cost consequences.
  \item \textbf{Issue Group:} Scheme of Arrangement
  \item \textbf{Issue:} Whether creditors are allowed to object to classification in an untimely manner
\end{itemize}

\smallskip
\noindent\textbf{Example 3}

\noindent\textit{Input:} In \textit{British Transport Commission v Gourley} [1956] AC 185 at 206, the House of Lords explained that in an action for personal injuries the damages are always divided into two main parts. First, there is special damage, which has to be specially pleaded and proved, consisting of out-of-pocket expenses and loss of earnings incurred down to the date of trial. Secondly, there is general damage which the law implies and is not specially pleaded, including compensation for pain and suffering and, if the injuries lead to continuing or permanent disability, compensation for loss of earning power in the future.

\noindent\textit{Output:}
\begin{itemize}[leftmargin=1.5em,itemsep=0pt]
  \item \textbf{Key Principles:} For personal injury actions, damages are divided into two main parts. Special damages must be specially pleaded and proved, consisting of out-of-pocket expenses and loss of earnings incurred down to the date of trial. General damages are implied by law and include compensation for pain and suffering and the like.
  \item \textbf{Issue Group:} Damages
  \item \textbf{Issue:} Whether the heads of claim were general or special damages
\end{itemize}

\smallskip
\noindent\textbf{Example 4}

\noindent\textit{Input:} A Review Committee is a ``sifting mechanism'' to weed out frivolous complaints operating within a tight statutory timeframe: s~85(6) and (8) of the LPA. Its function is to direct the Council to dismiss the matter if it is unanimously of the opinion that the complaint is ``frivolous, vexatious, misconceived or lacking in substance''. It is not the function of a Review Committee to go beyond the heads of complaint presented to it, and attempt to identify additional heads of complaint.

\noindent\textit{Output:}
\begin{itemize}[leftmargin=1.5em,itemsep=0pt]
  \item \textbf{Key Principles:} A Review Committee is a `sifting mechanism' to weed out frivolous complaints that operates within a tight statutory timeframe.
  \item \textbf{Issue Group:} Judicial Review
  \item \textbf{Issue:} Analysis of complaint and the Review Committee's report
\end{itemize}

\smallskip
\noindent\textbf{Example 5}

\noindent\textit{Input:} To rely on the partial defence of diminished responsibility, an offender bears the burden of proving three cumulative requirements: (a) that he was suffering from an abnormality of mind; (b) that the abnormality of mind arose from a condition of arrested or retarded development of mind, any inherent cause, or was induced by disease or injury; and (c) the abnormality of mind substantially impaired his mental responsibility for his acts and omissions. ``Abnormality of mind'' refers to a state of mind so different from that of ordinary human beings that the reasonable man would term it abnormal.

\noindent\textit{Output:}
\begin{itemize}[leftmargin=1.5em,itemsep=0pt]
  \item \textbf{Key Principles:} To rely on the partial defence of diminished responsibility, an offender bears the burden of proving three cumulative requirements: (a) suffering from an abnormality of mind; (b) that the abnormality arose from arrested development, inherent cause, or disease or injury; and (c) the abnormality substantially impaired his mental responsibility. Where the accused premeditates to kill under a veneer of rationality but the decision is the product of a disordered mind, the accused must show that but for the abnormality he would not have made that decision, and that in executing the intention he had no realistic moment of rationality enabling him to resile.
  \item \textbf{Issue Group:} Defence of diminished responsibility
  \item \textbf{Issue:} Whether the partial defence of diminished responsibility under Exception 7 to s~300 of the Penal Code is available to the offender
\end{itemize}

\smallskip
\noindent\textbf{Example 6}

\noindent\textit{Input:} A plaintiff is entitled to plead and claim on the basis of both the wrongful gain interest and the wrongful loss interest in an action for breach of confidence. The law is clear that an action for breach of duty of confidence in equity and an action for breach of contractual duty of confidentiality are distinct causes of action and should not be conflated.

\noindent\textit{Output:}
\begin{itemize}[leftmargin=1.5em,itemsep=0pt]
  \item \textbf{Key Principles:} A plaintiff is entitled to plead and claim on the basis of both the wrongful gain interest and the wrongful loss interest in an action for breach of confidence.
  \item \textbf{Issue Group:} Breach of duty of confidence
  \item \textbf{Issue:} Applicable law
\end{itemize}

\smallskip
\noindent\textbf{Example 7}

\noindent\textit{Input:} In \textit{Law Society of Singapore v Wong Sin Yee} [2018] 5 SLR 1261, the court found that by posing questions in cross-examination that were ``indecent, scandalous and calculated to insult or annoy'', the respondent solicitor's conduct was ``disgraceful and a clear abuse of the privileges of cross-examination''. Misconduct under the LPA is a broader catch-all provision where the conduct is deemed unacceptable but does not fall within any other enumerated ground. The standard applied is whether reasonable people, on hearing what the solicitor had done, would have said without hesitation that as a solicitor he should not have done it.

\noindent\textit{Output:}
\begin{itemize}[leftmargin=1.5em,itemsep=0pt]
  \item \textbf{Key Principles:} By posing questions in cross-examination that were ``indecent, scandalous and calculated to insult or annoy'', the respondent solicitor's conduct was disgraceful and a clear abuse of the privileges of cross-examination. Misconduct under the LPA is a broader catch-all provision where the conduct does not fall within any other enumerated ground. The standard is whether reasonable people, on hearing what the solicitor had done, would have said without hesitation that as a solicitor he should not have done it.
  \item \textbf{Issue Group:} Disciplinary Proceedings
  \item \textbf{Issue:} Whether the respondent's conduct constitutes grossly improper conduct as a form of misconduct under the LPA
\end{itemize}

\smallskip
\noindent\textbf{Example 8}

\noindent\textit{Input:} The English authorities make clear that the purpose of the statutory filter requiring authorisation for charity proceedings is to prevent the wasteful expenditure of precious charity funds on litigation. In \textit{Dean v Burne} [2009] EWHC 1250 (Ch), Blackburne J took the view that s~33(4) did not obviate the need for authorisation where the counterclaim was in the nature of a wholly distinct claim and in no sense arose out of the subject matter of the action.

\noindent\textit{Output:}
\begin{itemize}[leftmargin=1.5em,itemsep=0pt]
  \item \textbf{Key Principles:} Section 33(4) did not obviate the need for authorisation where the counterclaim was in the nature of a wholly distinct claim and in no sense arose out of the subject matter of the action.
  \item \textbf{Issue Group:} Statutory interpretation
  \item \textbf{Issue:} Statutory interpretation on authorisation
\end{itemize}

\smallskip
\noindent\textbf{Example 9}

\noindent\textit{Input:} The implied undertaking arises from the principle established in \textit{Riddick v Thames Board Mills Ltd} [1977] QB 881. Where a party to litigation has been ordered to give discovery, the discovering party may not use the discovered documents, and the information obtained therefrom, for a purpose other than pursuing the action in respect of which discovery is obtained. Public interest requires full and complete disclosure in the interest of justice, but discovery on compulsion is also an intrusion of privacy. The Riddick principle seeks to strike a balance between these two interests.

\noindent\textit{Output:}
\begin{itemize}[leftmargin=1.5em,itemsep=0pt]
  \item \textbf{Key Principles:} Where a party to litigation has been ordered to give discovery, the discovering party may not use the discovered documents, and the information obtained therefrom, for a purpose other than pursuing the action in respect of which discovery is obtained. The Riddick principle seeks to strike a balance between the public interest in full disclosure and the intrusion of privacy caused by compelled discovery.
  \item \textbf{Issue Group:} Whether documents can be adduced as further evidence on appeal
  \item \textbf{Issue:} The Riddick principle
\end{itemize}

\smallskip
\noindent\textbf{Example 10}

\noindent\textit{Input:} In \textit{Yien Yieh Commercial Bank Ltd v Kwai Chung Cold Storage Co Ltd} [1989] 2 HKLR 639, Lord Goff stated that to reject one clause in a contract as inconsistent with another involves a rewriting of the contract which can only be justified where the two clauses are in truth irreconcilable. Where the document has been drafted as a coherent whole, repugnancy is extremely unlikely to occur. The contract has to be read as a whole, and the overwhelming probability is that an apparent inconsistency will be resolved by the ordinary processes of construction.

\noindent\textit{Output:}
\begin{itemize}[leftmargin=1.5em,itemsep=0pt]
  \item \textbf{Key Principles:} To reject one clause in a contract as inconsistent with another involves a rewriting of the contract which can only be justified where the two clauses are in truth irreconcilable. Where the document has been drafted as a coherent whole, repugnancy is extremely unlikely to occur, and an apparent inconsistency will ordinarily be resolved by the ordinary processes of construction.
  \item \textbf{Issue Group:} Consistency between enforcement procedure and notice provisions
  \item \textbf{Issue:} Legal principles on whether the bank has a right to unwind
\end{itemize}

\smallskip
\noindent\textbf{Example 11}

\noindent\textit{Input:} A quantitative assessment of the relative number of syllables which two marks have in common is usually conducted in ascertaining aural similarity. Allowances should be made for imperfect recollection and careless pronunciation. Ultimately, it is the pronunciation of the relevant words as a whole that must be critical in ascertaining aural similarity.

\noindent\textit{Output:}
\begin{itemize}[leftmargin=1.5em,itemsep=0pt]
  \item \textbf{Key Principles:} It is the pronunciation of the relevant words as a whole that must be critical in ascertaining aural similarity.
  \item \textbf{Issue Group:} Trademark infringement
  \item \textbf{Issue:} Similarity between trademarks --- aural similarity
\end{itemize}

\smallskip
\noindent\textbf{Example 12}

\noindent\textit{Input:} The courts have indicated their willingness to imply a term of mutual trust and confidence where warranted. In \textit{Johnson v Unisys Ltd} [2003] 1 AC 518, Lord Hoffmann observed that over the last 30 years the nature of the contract of employment has been transformed, and that a person's employment gives not only a livelihood but an occupation, an identity and a sense of self-esteem. The contribution of the common law has been by the evolution of implied terms in the contract of employment, the most far-reaching of which is the implied term of trust and confidence.

\noindent\textit{Output:}
\begin{itemize}[leftmargin=1.5em,itemsep=0pt]
  \item \textbf{Key Principles:} The courts have indicated their willingness to imply a term of mutual trust and confidence where warranted. The most far-reaching contribution of the common law to employment contracts has been the implied term of trust and confidence.
  \item \textbf{Issue Group:} Bonuses
  \item \textbf{Issue:} Mutual trust and confidence to determine entitlement to bonuses
\end{itemize}

\smallskip
\noindent\textbf{Example 13}

\noindent\textit{Input:} The elements of the tort of unlawful means conspiracy can be summarised as follows: (a) there must be a combination of two or more persons and an agreement between or among them to do certain acts; (b) the conspiracy must be done by ``unlawful means'', which covers both a criminal act as well as an intentional tortious act; (c) the act must actually be performed in furtherance of the agreement; (d) there must be an intention to injure the plaintiff; and (e) damage must be suffered by the plaintiff.

\noindent\textit{Output:}
\begin{itemize}[leftmargin=1.5em,itemsep=0pt]
  \item \textbf{Key Principles:} The elements of the tort of unlawful means conspiracy include: (a) a combination of two or more persons and an agreement to do certain acts; (b) the conspiracy must be done by unlawful means; (c) the act must be performed in furtherance of the agreement; and (e) damage must be suffered by the plaintiff.
  \item \textbf{Issue Group:} Tort of unlawful means conspiracy
  \item \textbf{Issue:} Whether there is a claim under the tort of unlawful means conspiracy
\end{itemize}

\smallskip
\noindent\textbf{Example 14}

\noindent\textit{Input:} In \textit{Digicel (St Lucia) Ltd v Cable and Wireless plc} [2008] EWHC 2522 (Ch), Justice Morgan observed that what is generally required by an order for standard disclosure is a ``reasonable search'' for relevant documents, and that the rules do not require that no stone should be left unturned. As Lord Justice Jacob observed in \textit{Nichia Corporation v Argos Limited} [2007] EWCA Civ 741: ``Perfect justice'' in one sense involves a tribunal examining every conceivable aspect of a dispute, but a system which sought such perfect justice in every case would actually defeat justice. A compromise is made: one makes do with a lesser procedure even though it may result in justice being rougher. Better justice is achieved by risking a little bit of injustice.

\noindent\textit{Output:}
\begin{itemize}[leftmargin=1.5em,itemsep=0pt]
  \item \textbf{Key Principles:} Standard disclosure requires only a ``reasonable search'' for relevant documents. A system seeking perfect justice in every case would defeat justice due to disproportionate cost and time. Better justice is achieved by accepting a lesser procedure, even at the risk of some roughness --- better justice is achieved by risking a little bit of injustice.
  \item \textbf{Issue Group:} Electronic Discovery
  \item \textbf{Issue:} Extent to which inspection of compound documents is allowed
\end{itemize}

\smallskip
\noindent\textbf{Example 15}

\noindent\textit{Input:} In \textit{Re MC Bacon Ltd} [1990] BCLC 324, Millett J discussed the changes brought by the UK Insolvency Act 1986 in relation to unfair preference. It is no longer necessary to establish a dominant intention to prefer; it is sufficient that the decision was influenced by the requisite desire. It is no longer sufficient to establish an intention to prefer --- there must be a desire to produce the effect mentioned in the subsection. It is the subjective desire of the company to improve the creditor's position in the event of its imminent insolvency that is decisive. There is no need for direct evidence of the requisite desire as it can be inferred from the circumstances of the case.

\noindent\textit{Output:}
\begin{itemize}[leftmargin=1.5em,itemsep=0pt]
  \item \textbf{Key Principles:} It is no longer necessary to establish a dominant intention to prefer; it is sufficient that the decision was influenced by the requisite desire. There must be a desire to produce the effect mentioned in the subsection, not merely an intention to prefer.
  \item \textbf{Issue Group:} Unfair preference
  \item \textbf{Issue:} Whether the plaintiff was influenced by a desire to prefer the defendant
\end{itemize}

\section{\textcolor{black}{Cold-Start Principle Prediction Prompt}}\label{app:coldstart-prompt}

\textcolor{black}{The following prompt was used for cold-start principle prediction (Section~\ref{sec:coldstart}). DeepSeek-V3 was instructed to predict the most likely legal principle from a fact summary alone, with case names disallowed in the output to prevent leakage of citation identity into the retrieval query.}

\smallskip
\noindent\textcolor{black}{\textbf{System prompt:} \textit{You are an expert legal analyst assisting a Singapore lawyer who needs to find precedent cases. The lawyer has described their client's situation but has not yet identified the specific legal authority they need.}}

\smallskip
\noindent\textcolor{black}{\textbf{Task instruction:} Given a factual description of a case, predict the single most likely legal principle the lawyer would need precedent authority for. Output the principle in full formal legal phrasing as it might appear in a Singapore judgment.}

\smallskip
\noindent\textcolor{black}{\textbf{Rules:}}
\textcolor{black}{
\begin{itemize}[leftmargin=1.5em,itemsep=0pt]
  \item Output exactly ONE legal principle.
  \item Use formal legal phrasing (e.g., ``The duty of care arises when\ldots'', ``An adverse inference may be drawn where\ldots'').
  \item Focus on the substantive legal proposition, not procedural rules unless procedure is the central issue.
  \item Do not name specific cases or cite case names.
  \item Do not include preamble such as ``The principle is\ldots'' or ``Based on these facts\ldots''.
  \item Output the principle directly as a self-contained statement.
\end{itemize}
}

\smallskip
\noindent\textcolor{black}{\textbf{Output format:} A single sentence or short paragraph stating the legal principle. No additional text.}

\smallskip
\noindent\textcolor{black}{\textbf{Input variable:} \texttt{\{fact\_summary\}} is replaced with the LLM-generated fact summary from Section~\ref{sec:dataset} (Step 1).}

\smallskip
\noindent\textcolor{black}{\textbf{Contamination safeguards.} Before passing fact summaries to DeepSeek-V3, we screened them for Singapore neutral citation patterns (e.g., \texttt{[2024] SGHC 123}), ``Party v Party'' patterns, and citation keywords (e.g., SGCA, SGHC, SLR, MLJ); the same screen was applied to predicted principles. Across the 50 sampled test pools, no fact summary and no predicted principle was flagged.}

\section{Data Consent, Recruitment And Payment}\label{app:data-construction}
We engaged legal experts to annotate, e.g., case names, legal principles, issues, and issue groups based on citing judgments. All annotators were informed in advance that their contributions would be used solely for academic research purposes and provided informed consent prior to participation. We do not collect or disclose any sensitive personal information, including demographic details or private data of the annotators. All data handling procedures adhere to standard research ethics and privacy guidelines to ensure confidentiality and responsible use.

\definecolor{MyColor1}{RGB}{125,171,186}
\definecolor{MyColor2}{RGB}{174,188,207}
\definecolor{MyColor3}{RGB}{239,170,72}
\definecolor{MyColor4}{RGB}{173,196,178}
\definecolor{MyColor5}{RGB}{198,136,171}
\definecolor{MyColor6}{RGB}{226,169,241}
\definecolor{MyColor7}{RGB}{203, 108, 230}
\definecolor{MyColor8}{RGB}{0, 191, 99}

\begin{table*}[!htbp]
  \vspace{-0.1in}
  \scriptsize
  \begin{tabular}{p{0.47\textwidth}||p{0.47\textwidth}}
    \toprule
    \cellcolor{MyColor1}\textbf{Citing Judgment} &\cellcolor{MyColor3}\textbf{Issue} 
    \\
    \textit{Re Ariffin Iskandar Sha bin Ali Akbar and other matters} [2025] SGHC 156 &Whether the stakeholders agree that Mr Foo should be admitted to the Bar.  
    \\\hline
    \cellcolor{MyColor6}\textbf{Raw Fact} (scraped Background section, 1,139 words, abbreviated) &\cellcolor{MyColor8}\textbf{Citation Paragraph} (987 words, abbreviated) \\
    Where the court determines that an applicant is not yet suitable for 
    admission to the Bar on account of some issue of character, the usual 
    course has been to invite the applicant to withdraw his or her 
    application [...] The applications before me are three such Legacy 
    Cases, which were due for consideration just as the Legal Profession 
    (Admission) Rules 2024 took effect [...] while the applicants might 
    not yet be fit for admission, withdrawal would not be appropriate. 
    & The second incident occurred while Mr Foo was taking the LAW204 
    Constitutional \& Administrative Law module [...] However, his essay 
    contained phrases that appeared to have been lifted from Wikipedia 
    without attribution [...] The AG, SILE and LSS accepted that the 
    LAW204 Incident did not disclose dishonesty, but a lack of academic 
    diligence: \textit{Re Suria Shaik Aziz} [2023] 5 SLR 1272 at [25]. 
    They therefore submitted that the LAW204 Incident did not affect 
    Mr Foo's suitability of character. [...] Mr Foo did not contest the 
    position taken by the AG and the SILE in respect of deferring his 
    application. 
    \\\hline
    \cellcolor{MyColor7}\textbf{Fact} (LLM-summary) &\cellcolor{MyColor5}\textbf{Key Principle Illustrated}\\
    %\midrule
    My client was found not yet fit for admission to the Bar due to a 
    character issue, but his application was stayed for 18 months instead 
    of dismissed because he qualified under the old admission rules. The 
    stay allows him time for rehabilitation without forcing him to retake 
    exams and restart his training period. 
    & Dishonesty is to be distinguished from a lack of academic diligence. \\\hline
    \cellcolor{MyColor2}\textbf{Issue Group} &\cellcolor{MyColor4}\textbf{Cited Case}\\
    %\midrule
    Admission of Candidate &\textit{Re Suria Shaik Aziz} [2023] 5 SLR 1272 
    \\\bottomrule
  \end{tabular}
  \vspace{-0.1in}
  \caption{An example of SG-LegalCite extraction.}
  \label{tab:extraction_example}
  \vspace{-0.1in}
\end{table*}

The three legal expert annotators were recruited in the expert validation process. Each annotator was compensated SGD 100 per grading round, with two rounds conducted in total — one for evaluating DeepSeek-V3 extractions and one for evaluating Claude Sonnet 4 extractions. The legal expert who performed the initial annotations for LLM selection (25 judgments, 725 case-principle pairs) was compensated SGD 10 per judgment annotated.

Specifically, the three experts independently rated 131 case-principle pairs from 15 randomly sampled judgments using a 5-point scale across three dimensions on the key principle illustrated, issue and issue group. 
\begin{itemize}[leftmargin=*, noitemsep]
    \item \emph{Key Principles Illustrated}, assessed on whether the legal principles are accurately extracted from the cited paragraph, represent actual holdings or dicta, and are relevant to the citing context; 
    \item \emph{Issue}, assessed on whether it correctly identifies the legal question addressed, is sufficiently specific, and matches the citation context;
    \item \emph{Issue Group}, \textcolor{black}{a fine-grained doctrinal tag based on the issue, assessed on whether it accurately captures the doctrinal area of the citation and is consistent with the citation context.}
\end{itemize}

\section{Domain Distribution}\label{app:domain-distribution}

\begin{figure}[ht]
    \centering
    \includegraphics[width=\columnwidth]{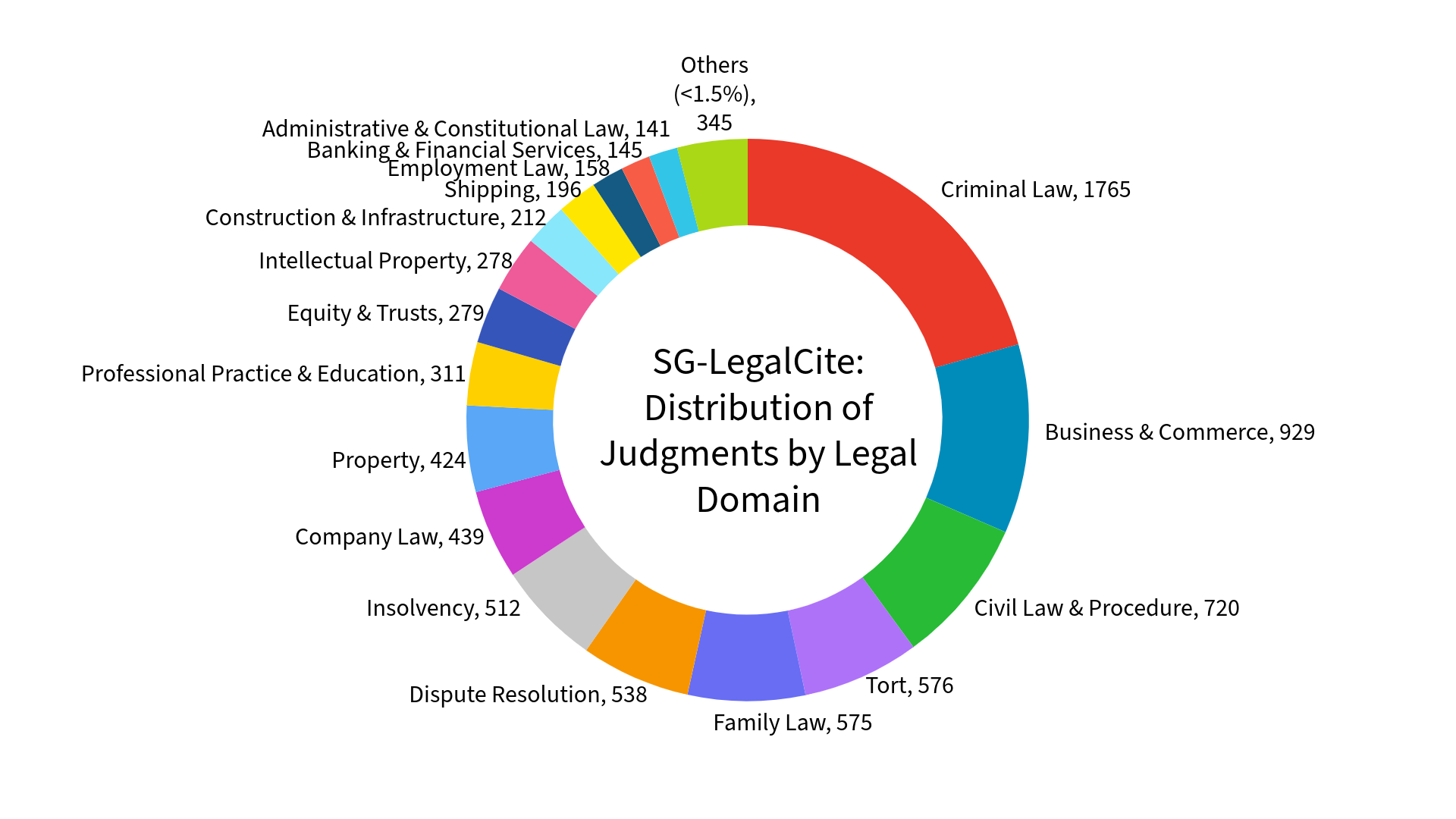}
    \caption{Distribution of judgments by legal domain in SG-LegalCite.}
    \label{fig:domain-distribution}
\end{figure}

\textcolor{black}{Figure~\ref{fig:domain-distribution} shows the distribution of the 8,523 judgments in SG-LegalCite across legal domains, classified using the 34 practice-area tags used by Singapore Law Watch~\cite{singaporelawwatch}. Domains with fewer than 1.5\% of the corpus are grouped as ``Others (<1.5\%)''.}

\section{An Example of Citation-level Record}\label{app:example-record}
Table~\ref{tab:extraction_example} provides a concrete example of a citation-level record produced by our pipeline, illustrating how raw factual and citation context are transformed into the structured fields in SG-LegalCite.

\section{Detailed Settings of Doctrinal Signal Granularity Ablation}\label{app-granularity-ablation}
\textcolor{black}{For a fair comparison, all models share identical candidate pools, negative samples, and hyperparameters: batch size = 64; learning rate = 2e-5, 10 epochs with early stopping, QLoRA r=16 for SaulLM-7B; and only the query text differs.}

\section{Detailed Settings of Citation Context Extraction}\label{app:settings-citation-context}
\textcolor{black}{For each case--principle pair, we construct a \emph{scrubbed context window}: a $\pm$200-word window centred on the cited case mention in the original citation paragraph, from which all case names, neutral citations, reporter references, judge names, and pinpoint markers are removed to prevent retrieval via surface matching ({residual identifiability $<$0.05\% of cases}). To ensure a fair comparison, the scrubbed context is then truncated to match the word count of the corresponding principle for that row ($\sim$ 70 tokens on average), so both query types receive the same text budget.}

\textcolor{black}{Regarding the residual identifiability, it is calculated in the following way. A multi-layer regex pipeline was applied to scrub the citation context windows. The pipeline removes: (1) the target cited case name; (2) all other case names in the paragraph, covering six patterns: Party v Party, PP v X, R v X, Re X, Ex parte X, and In the Matter of X; (3) Singapore neutral citations (e.g., [2024] SGCA 5); (4) reporter references (SLR, MLJ, AC, WLR, KB, etc.), including volume-reporter-page format (e.g., ``1 KB 1''); (5) judge names with titles (e.g., Lord Denning, Phang JA, Stamp LJ); (6) bare year markers (e.g., [1985], (1848)); and (7) pinpoint references (e.g., at [60], at paras 14--15). Each match is replaced with a placeholder token ([CASE], [CITATION], [JUDGE], [YEAR], or [REPORTER]) to preserve sentence structure while removing identifiers. The scrubber was developed over four iterations (V1--V4), with a full-dataset audit after each round. V1 exhibited approximately 13\% residual leakage, primarily from case names with single-letter parties (e.g., ``R v S'') and lowercase prefixes (e.g., ``de Dampierre v de Dampierre'') that the initial regex patterns did not capture. V4 addressed these by allowing single-character party names and lowercase prefix tokens (de, von, van, etc.) in the case-name regex. To verify the final scrubbing quality, we audited all 100,890 rows (not a sample) against 18 leakage-detection patterns. 13 of 18 patterns produced zero hits. The remaining five patterns yielded a combined total of approximately 50 rows with any surviving identifiable reference (22 real case-name leaks from unusual company-name formatting, plus 28 minor residuals such as bare ``ex parte'' phrases and stray year tokens), giving a residual identifiability rate of approximately 0.05\%.}

\end{document}